\documentclass[%
 reprint,
 amsmath,amssymb,
 aps,
]{revtex4-1}

\usepackage{graphicx}
\usepackage{epstopdf}
\usepackage{dcolumn}
\usepackage{bm}

\begin{document}


\title{A Quantum Theory of Angle and Relative Phase Measurement} 



\author{Scott Roger Shepard}
\email{sshepard@latech.edu}

\affiliation{College of Engineering and Science, Louisiana Tech University, Ruston LA 71272.}



\begin{abstract}
The complementarity between time and energy, as well as between an angle and a component of angular momentum, is described at three different layers of understanding. The first layer is comprised of a simple Fourier transform of the complementary wavefunction and we elucidate ways in which these can be interpreted as fuzzy measurements. The limited dimensionality of the state space is shown to prevent wavefunction collapse. Therein an auxiliary system is shown to function as a noise source and the fuzzy statistics manifest. We go beyond these and define a sharp (collapsible) measurement of relative phase by exciting the auxiliary mode on an important subspace of the more general two-mode state space; which yields the second layer of understanding. The phenomena of super-resolution are readily apparent in the quantum phase representation which also reveals that entanglement is not required. We modify Schwinger's harmonic oscillator model of angular momentum to include the case of photons. Therein the quantum angle measurement is shown to be equivalent to the relative phase measurement between the two oscillators. The meaning of relative phase is finalized at the third layer of understanding on an unrestricted two-mode space which reduces an infinite number of possible measurements to two reasonable ways of dealing with degeneracy. These also correspond to two reasonable ways of eliminating absolute time in order to define a direct measurement of the relative phase: a conditional measurement which takes a snapshot in absolute time (corresponding to adding probability amplitudes); and a marginal measurement which takes an average in absolute time (corresponding to adding probabilities). The sense in which distinguishability is a ``matter of how long we look'' is discussed and the meaning of the general theory is illustrated by taking the two oscillators to be photons so that the conditional measurement reveals a snapshot of the angular distribution of the electric field vector and the marginal measurement corresponds to a quantum version of the polarization ellipse. It is shown that an odd number of x-polarized photons will \textit{never} have an angle in correspondence with the y-axis; but an even number of x-polarized photons \textit{always} can! The behavior of an x-polarized coherent state is examined and the snapshot angular distributions are seen to evolve into two counter-rotating peaks resulting in considerable correspondence with the y-axis  at the time for which a classical linear polarization vector would shrink to zero length. We also demonstrate how the probability distribution of absolute time (now a measurable quantity, rather than just a parameter) has an influence on how these snapshot angular distributions evolve into the quantum polarization ellipse.
\begin{description}
\item[PACS numbers]
03.65.Ta, 42.50.Tx, 42.50.St, 42.50.Lc, 42.50.Dv, 42.50.Ex
\end{description}
\end{abstract}


\maketitle 

\section{Single Oscillator Phase Statistics}
\subsection{Introduction}

Dirac [1], in 1927, postulated the existence of an Hermitian phase operator, $\hat{\phi}$, as part of a polar decomposition of the annihilation operator, $\hat{a}=e^{i\hat{\phi}}(\hat{a}^\dagger\hat{a})^{1/2}$, for a single quantum harmonic oscillator, where $\hat{a}^\dagger\hat{a} = \hat{n}$ is the photon number operator, the eigenvalues of which correspond to the energy levels of the oscillator. In this ordering the operator $e^{i\hat{\phi}}=\hat{a}(\hat{n})^{-1/2}$  lowers the photon number (like $\hat{a}$ but without any $\sqrt{n}$ factor)  
\begin{equation}
{e^{i\hat{\phi}}}|n\rangle = |n-1\rangle \ (\forall n \ge1)
\end{equation}
however, the action on the vacuum state is undefined. In 1964, Susskind and Glogower [2] demonstrated that no such Hermitian operator exists on the denumerably infinite dimensional  space [3] spanned by the number-kets $\{|n\rangle:n=0,1,2, . . . \ \infty \}$. They proposed, instead, the use of a polar decomposition of $\hat{a}$ with the opposite ordering; the  Susskind-Glogower (SG) phase operator, $\widehat{e^{i\phi}}\equiv(\hat{a}\hat{a}^\dagger)^{-1/2}\hat{a}$, so that the SG operator is a pure lowering operator which stops (i.e., yields the null ket)  at the vacuum:
\begin{equation}
\hat{A}|n\rangle = |n-1\rangle \ (\forall n\ge1)\ \ \text{and} \ \ \hat{A}|0\rangle = 0,
\end{equation}
where to simplify the notation we let $ \hat{A} \equiv \widehat{e^{i \phi}}$. Thus the SG operator has a number representation given by: $\hat{A} = \sum_{n=0}^{\infty} | n \rangle \langle n +1 |$. Due to the bound on photon number eigenspectra (i.e., the absence of negative energy eigenstates) this translation however cannot be unitary. It can only be one-sided unitary, i.e., $\hat{A} {\hat{A}}^{\dagger} = \hat{I}$ but ${\hat{A}}^{\dagger}\hat{A} = \hat{I} - \hat{V}$ where $\hat{I}$ is the identity operator on the state space of a single harmonic oscillator and $\hat{V} \equiv |0\rangle \langle 0 |$ is the vacuum projector. Thus, pure translation on a bounded state space cannot be unitary; therefore it cannot be expressed as the power series of an Hermitian (i.e., self-adjoint) operator. 

In light of an ever improving understanding of what it means to associate a measurement with an operator which doesn't commute with its adjoint, we [4], [5] demonstrated a connection between the SG operator and Helstrom's maximum likelihood (ML) quantum phase estimator [6]. Helstrom was not concerned with polar decompositions of the annihilation operator, nor with obtaining a description of a phase measurement that is complementary to that of photon counting. 
The ML measurement is based on the state-dependent kets: 
\begin{equation}
|\phi,\psi \rangle = \sum_{n=0}^{\infty} e^{i (n \phi + \chi_{n})} |n \rangle
\end{equation}
where the $ \chi_{n}$ are the phases of the number-ket expansion coefficients,  $\psi_n \equiv \langle n | \psi \rangle$.
Thus, the ML phase estimation procedure can be decomposed into two steps: the first  being to effectively remove the phases of the input quantum state's number-ket expansion coefficients,  $\psi_n$, i.e., these phases are all effectively set to be equal to zero. This first step can only be omitted if the $\psi_n$ are already real.  The second step turns out to be equivalent to performing a relative phase measurement between two oscillators  (complementary to the measurement of the difference in their photon numbers) when one of those oscillators is in the vacuum state [7], [8] --- resulting in the single oscillator (or single-mode) phase statistics, which are $\it{fuzzy}$ in a sense that we will elucidate herein. Thus the ML statistics correspond to the single-mode statistics for the case of states with  real $\psi_n$, as depicted in the Venn diagram of Fig.\ 1. 

Concurrently (with respect to [4], [5]) and independently, an alternate method for obtaining the single oscillator statistics was derived by Pegg and Barnett  [9], [10]. Their approach requires the truncation of the infinite-dimensional state space of a harmonic oscillator to one of finite but arbitrarily large dimension. This subspace, denoted ${\mathcal{H}^T}(s)$, is spanned by the number-kets $\{ | n \rangle : 0 \le n \le s \}$. When $s$ is finite the resulting Pegg-Barnett (PB) discrete-phase kets are an orthogonal subset of the single-mode continuous phase-kets, as depicted in Fig.\ 1. This approach is described in Appendix 1 but it is important to note in passing that: {\it when s is finite} the discrete-phase measurement is ``sharp'' (i.e., it permits wavefunction collapse via projections onto the eigenkets of an Hermitian operator). Yet we also have that:  {\it when instead} the limit $s \rightarrow \infty$ is taken, 
these discrete-phase statistics converge (in distribution, i.e., in as much as rational numbers can converge to real numbers) to those of  a fundamentally ``fuzzy'' measurement --- the single-mode continuous phase statistics (i.e., the continuous envelope to which they converge cannot collapse to a delta-function). Thus, unless $s$ is left to be finite, this is an alternate means of calculating the same statistics via a limiting procedure. 

\begin{figure}
\centering 
\includegraphics[scale=.4]{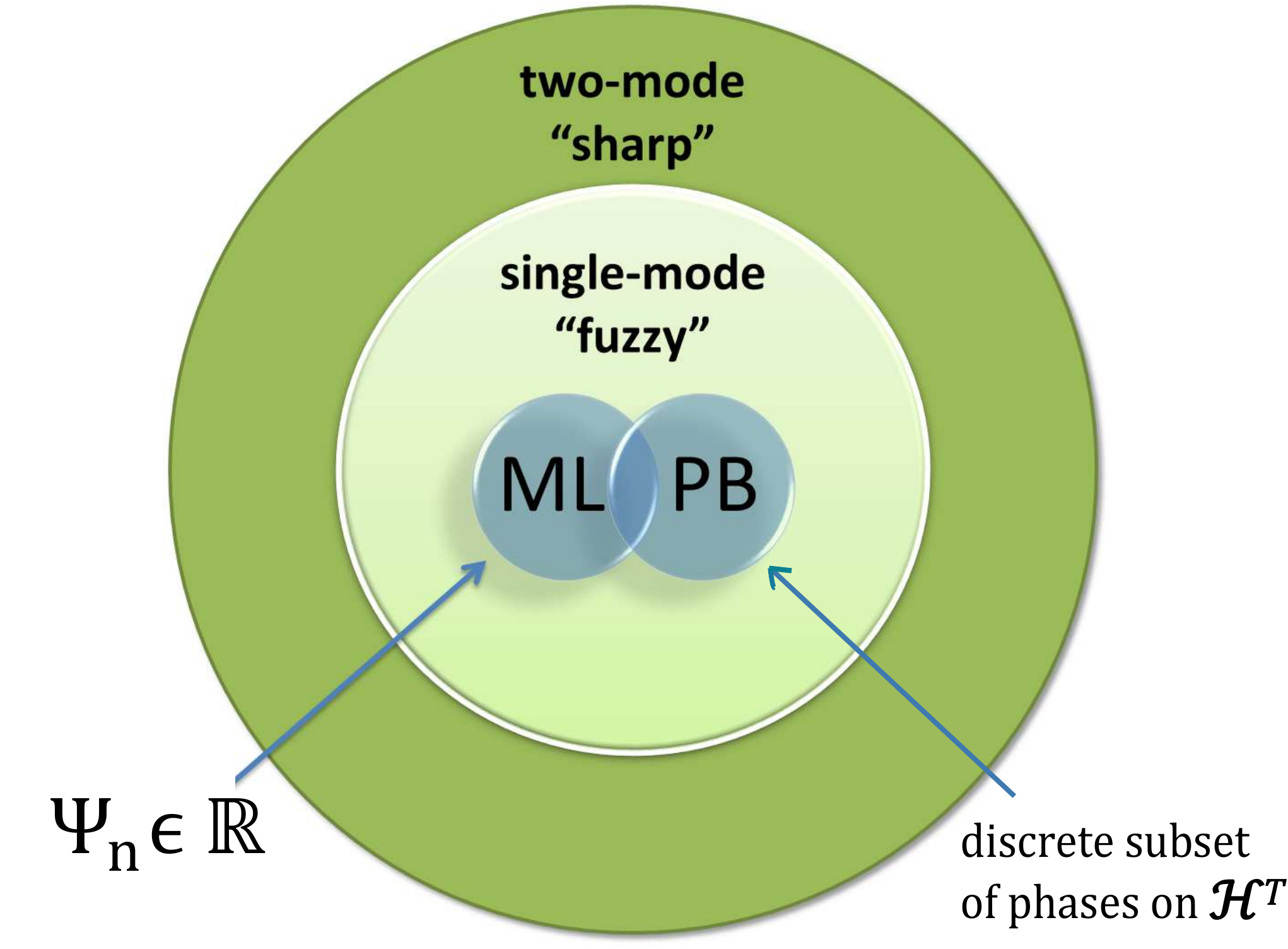}
\caption{Hierarchy of Some Quantum Phase Measurements}
\end{figure}

The connection to the SG operator is as follows. Although the SG operator has other eigenkets (the coherent phase states [11]) we  restrict our attention to the infinite energy subset of these --- the continuous phase-kets, of number representation: 
\begin{equation}
 |\phi \rangle = \sum_{n=0}^{\infty} e^{in\phi} | n \rangle
\end{equation}
which are normailzable in the continuous $\phi$ domain [12]. The restriction to these was made not only to connect with the essence of ML phase estimation; but (more significantly) to obtain complementarity to the measurement of photon number. These kets have been said [13] to be of fundamental importance because they underlie ML phase estimation. It was also noted [7], [8] however that these phase-kets are of even greater significance because they provide the first of three layers of understanding complementarity in a more general context. Complementarity at the first layer corresponds to the fuzzy single-mode case (depicted in Fig. 1, which generalizes the ML measurement and to which the PB measurement can converge). 
The second and third layers of understanding  can only be achieved when we extend beyond the fuzzy measurement by working in a state space larger than that of a single oscillator --- as we will in the next section. 

The SG operator does not commute with its adjoint, hence it is not comprised of a set of commuting Hermitian operators. Therefore the measurement statistics we associate with it cannot be calculated via the familiar Hermitian operator rules (e.g.,  moments calculated via $\langle\psi|\hat{A}^{k}|\psi\rangle$ for $k=1, 2,...$ do \textit{not} correspond to the single-mode statistics).

 Quantum measurements can be described in terms of wavefunctions (as well as operators) and perhaps the simplest path to the single-mode statistics is to form the phase wavefunction
\begin{equation}
\psi(\phi) \equiv \langle\phi|\psi\rangle,
\end{equation} 
from which the phase probability distribution $P(\phi)=|\psi(\phi)|^{2}/(2 \pi)$, and its associated moments follow directly. This will be formally justified below but it stems from the fact that the (non-orthogonal) phase-kets are complete: 
\begin{equation}
 \int_{-\pi}^{\pi} \frac{d\phi}{2\pi} |\phi \rangle \langle \phi | = \sum_{n=0}^{\infty} |n \rangle \langle n | = \hat{I}.
\end{equation}
 
The resolution of the identity by the phase-kets in (6) permits the extremely useful phase representation of an arbitrary quantum state:
\begin{equation}
| \psi \rangle = \hat{I} | \psi \rangle = \int_{-\pi}^{\pi} \frac{d\phi}{2\pi} \langle \phi | \psi \rangle \ | \phi \rangle,
\end{equation}
analogous to the familiar number-ket expansion of a state:
\begin{equation}
| \psi \rangle = \hat{I} | \psi \rangle = \sum_{n=0}^{\infty} \langle n | \psi \rangle \ | n \rangle.
\end{equation}
Just as the number-ket expansion coefficients, $\psi_n \equiv \langle n | \psi \rangle$, may be viewed as a wavefunction in discrete n-space, the inner product $\psi(\phi) \equiv \langle \phi | \psi \rangle$ is a wavefunction in continuous $\phi$--space. 
The Fourier transform between the  number and phase wavefunctions:
\begin{equation}
\psi(\phi) = \sum_{n=0}^{\infty} \psi_n e^{-in\phi} \ \longleftrightarrow \ \psi_n = \int_{-\pi}^{\pi} \frac{d\phi}{2\pi} \psi(\phi) e^{in\phi},
\end{equation}
demonstrates the complementarity of photon number and quantum phase. 

Thus, to reveal the phase properties of an arbitrary state, we can simply take the familiar number-ket expansion coefficients, $\psi_n$, to be the Fourier series coefficients for $\psi(\phi)$ --- which underlies the harmonic oscillator's continuous and periodic phase distribution. Several relations among $\psi_n$ and $\psi(\phi)$ are reminiscent of those encountered in Schrodinger's wave mechanics. Analogous to the position representation of the momentum operator $\hat{p} \rightarrow -i\hbar \frac{d}{dx}$, for example, we have a phase representation of the number operator, $\hat{n} \rightarrow - i\frac{d}{dx}$:

\begin{eqnarray}
\langle (\hat{n})^k \rangle = \int_{-\pi}^{\pi} \frac{d\phi}{2\pi} \psi^* (\phi) (-i \frac{d}{d\phi})^k \psi(\phi) \nonumber \\ (  \forall \ \rm{integer} \ \it{ k}).
\end{eqnarray}
We note in passing, however, that differentiation with respect to a discrete variable is an undefined operation.

Alternatively we might have arrived at this, (4) -- (10), without any reference to the SG operator, via a general theory of complementarity (as is done in Appendix 2, and as is $\it{required}$ for the fuzzy description of the angle of a particle of finite angular momentum, since then there is also an upper bound on the complementary eigenspectra and that prohibits a lowering operator from having any eigenkets at all). Historically such an obvious approach (resulting in a Fourier transform between complementary wavefunctions) has been slow to gain acceptance ---  leading to a lack of time and angle operators in quantum theory --- due to the mathematical subtleties which arise from our perfectly valid predelection for Hermitian operators (and/or equivalently for wavefunctions that can collapse). One might argue that these can now be accepted as ``fuzzy'' measurements, i.e., non-projection valued POMs (probability operator measures) [14] so that, at layer 1, this $\it{is}$ a solution since these measurements do exist. The author however views  fuzzy measurements as ``incomplete descriptions'' [7], [8] of a realizable measurement; and would argue that one still needs to achieve a (sharp) ``complete description'' of such measurements in terms of sets of commuting Hermitian operators and their associated collapsible wavefunctions to fully understand the complementarity alluded to at this first layer.

A non-projection valued POM is simply a resolution of the identity operator by non-orthogonal eigenkets [14]. The bound in 
their complementary eigenspectra 
 prevents orthogonality for the single-mode phase-kets of (4). Equivalently, (9) describes a one-sided Fourier series, but it would take a two-sided Fourier series to represent a delta-function. The  limited dimensionality of the space in one domain cannot support such sharp behavior in the complementary wavefunction.
%
 This limitation does more than prohibit delta-functions, it restricts the class of  single-mode phase statistics to those which must  satisfy a Paley-Weiner theorem [15] which can be expressed as 
\begin{equation}
\int_{-\pi}^{\pi} \frac{d\phi}{2\pi} | log | \psi(\phi)|| < \infty .
\end{equation}
This theorem demonstrates, for example, that $|\psi (\phi)|$ cannot vanish over an interval of non-zero width. The limited dimensionality of the underlying wavefunctions simply cannot support such sharp behavior.  Thus the single-mode phase distribution, $P(\phi)$, can vanish (equal zero) only at isolated points in $\phi$.

\subsection{Examples of Naimark's Extension Theorem}

The fact that non-projection valued POMs do correspond to realizable quantum measurements can be made more palatable via Naimark's Extension Theorem [14] which (to paraphrase) states that these correspond to the measurement of sets of commuting Hermitian operators defined on a larger state space when subsets of that larger space are not entangled with the state of the original space prior to the measurement. In this section we describe two examples from quantum optics in which the original space $\mathcal{H}_{s}$ is that of a single-mode (i.e., a single   harmonic oscillator --- the original system of interest); the larger space is the product space $\mathcal{H}_{s} \otimes \mathcal{H}_{a}$ of the original mode with some additional (a.k.a.\ auxiliary) mode on $\mathcal{H}_{a}$ which fundamentally $\it{must}$ be a part of the physical apparatus which realizes the quantum measurement; and the additional mode is ``off''  (\hspace{-1mm} i.e., placed in the vacuum state) prior to the measurement. It is important to note from the onset however that Naimark's extension theorem provides a means of describing a measurement which is $\it{still}$ fuzzy in the aforementioned senses. It does not provide a general means of extending to a sharp measurement (as we will in the next section) although it can sometimes give clues as to how that might proceed. 
 
%

Clearly the operator $\hat{o} = \hat{x} + i \hat{y}$, where $\hat{x}$ and $\hat{y}$  are the commuting (hence simultaneously measureable) $x$ and $y$ position operators, is measureable in the sharp sense (since it commutes with its adjoint [16]).
For an operator such as $\hat{a} = \hat{x} + i \hat{p}$ however, where $\hat{p}$ is the x-component of the momentum operator, we certainly cannot directly associate a sharp measurement since a perfectly precise simultaneous measurement of its real and imaginary components would constitute a violation of the uncertainty principle (i.e., $\hat{a}$ does not commute with its adjoint). We can however associate a quantum measurement with such an operator in a fuzzy sense: a simultaneous measurement of its real and imaginary components which is not perfectly precise in either. That such a measurement exists stems from the fact that the eigenkets of this annihilation operator (the coherent states $| \alpha  \rangle $  [17]) are complete, i.e., they resolve the identity operator; and completeness alone is sufficient to guarantee that $|\langle  \alpha | \psi \rangle | ^{2}$  is a perfectly valid PDF (probability distribution function) which must therefore, in some sense, describe a realizable quantum measurement. The fact that the coherent states are not orthogonal is a reflection of the fact that this is a fuzzy measurement.
\newline
\indent
Fundamental to the realizable measurement of any operator which does not commute with its adjoint is the existence of an auxiliary noise source.  Zero-point fluctuations [18] from this auxiliary mode prevent a  perfectly precise simultaneous measurement of the non-commuting real and imaginary parts of the original operator (so that the uncertainty principle is not violated). We can see this more clearly by extending the operator of interest to a larger space.
Any quantum measurement described by a non-projection valued POM on $\mathcal{H}_s$ 
can be represented by a collection of commuting observables on a larger Hilbert space. The utility of this representation lies in the identification of the aforementioned noise source. This auxiliary system is an integral part of the physical apparatus which realizes the quantum measurement.
\newline
\indent
Formally, the procedure is to find the Naimark extension, on $\mathcal{H} = \mathcal{H}_s \otimes \mathcal{H}_a$, of the desired POM on $\mathcal{H}_s$. For our purpose, this amounts to finding an operator on $\mathcal{H}$, say ${\widehat{E}}_{s\otimes a}$, which commutes with its adjoint such that its real and imaginary parts form a pair of commuting observables (a.k.a.\  Hermitian operators). Furthermore, we require that the measurement statistics of ${\widehat{E}}_{s\otimes a}$ reproduce those of the original operator on $\mathcal{H}_s$ (associated with the desired POM) when the auxiliary system is in some appropriate quantum state.
\newline
\indent
We consider now a  product space description of the measurement associated with the annihilation operator $\hat{a}_s$, for a single quantum harmonic oscillator, defined on the  space $\mathcal{H}_s$. When this oscillator is used to model a single mode of the electromagnetic field, this measurement can be realized by the heterodyne detection process and the non-commuting real and imaginary parts of $\hat{a}_s$, denoted $\hat{\chi}_s$ and $\hat{\rho}_s$ respectively, represent the in-phase and quadrature field components. It has been shown [19] that one extension of $\hat{a}_s$ onto $\mathcal{H} = \mathcal{H}_s \otimes \mathcal{H}_a$ is
\begin{equation}
\hat{y} \equiv \hat{a}_s \otimes \hat{I}_a + \hat{I}_s \otimes \hat{a}_a^\dagger,
\end{equation}
where $\hat{a}_a = \hat{\chi}_a + i\hat{\rho}_a$ is the annihilation operator for the auxiliary mode. Since operators on different spaces commute, we have $[\hat{y},\hat{y}^\dagger] = 0$ so that the real and imaginary parts of $\hat{y}$, denoted as $\widehat{X}$ and $\widehat{P}$, comprise a pair of commuting observables.

Notice that when the auxiliary mode is in the vacuum state the expected values of  $\hat{X}$ and $\hat{P}$,  in this realizable measurement of $\hat{a}_s$, are the same as those of two $distinct$ measurements of the non-commuting operators, $\hat{\chi}_s$ and $\hat{\rho}_s$ on two identically prepared systems :
\begin{equation}
\langle \widehat{X}\rangle_{s\otimes a}
= \ _{s}\langle \psi | \hat{\chi}_s | \psi \rangle_s + \ _{a}\langle 0 | \hat{\chi}_a | 0 \rangle_a =  \ _{s}\langle \psi | \hat{\chi}_s | \psi \rangle_s
\end{equation}
and
\begin{equation}
\langle \widehat{P} \rangle_{s\otimes a}
 = \ _{s}\langle \psi | \hat{\rho}_s | \psi \rangle_s -  \ _{a}\langle 0 | \hat{\rho}_a | 0 \rangle_a = \ _{s}\langle \psi | \hat{\rho}_s | \psi \rangle_s
\end{equation}
where we used $\widehat{X} \equiv (\hat{y} + \hat{y}^{\dagger})/2 = \hat{\chi}_s \otimes \hat{I}_a + \hat{I}_s \otimes \hat{\chi}_a$ and $\widehat{P} \equiv (\hat{y} - \hat{y}^{\dagger})/2i = \hat{\rho}_s \otimes \hat{I}_a - \hat{I}_s \otimes \hat{\rho}_a$.

For the second moments however, we find 
\begin{equation}
\langle \widehat{X}^2 \rangle_{s\otimes a}= \ _{s}\langle \psi | {\hat{\chi}_s}^2 | \psi \rangle_s + \ _{a}\langle 0 | {\hat{\chi}_a}^2 | 0 \rangle_a = \\
  \ _{s}\langle \psi | {\hat{\chi}_s}^2 | \psi \rangle_s + 1/4
\end{equation}
 and 
\begin{equation}
\langle \widehat{P}^2 \rangle_{s\otimes a} = \ _{s}\langle \psi | {\hat{\rho}_s}^2 | \psi \rangle_s + \ _{a}\langle 0 | {\hat{\rho}_a}^2 | 0 \rangle_a = \ _{s}\langle \psi | {\hat{\rho}_s}^2 | \psi \rangle_s + 1/4.
\end{equation}
The variance of a measurement of $\hat{\chi}_s$ on the original system (of state space $\mathcal{H}_s$) is $\langle\Delta {\hat{\chi}_s}^2\rangle \equiv \  _s\langle\psi|{\hat{\chi}_s}^2|\psi\rangle_s - (  _s\langle\psi|\hat{\chi}_s|\psi\rangle_s )^2$. The variance of a $separate$ or independent measurement of $\widehat{\rho}_s$, on an identically prepared system, is $\langle\Delta {\hat{\rho}_s}^2\rangle \equiv \ _s\langle\psi|{\hat{\rho}_s}^2|\psi\rangle_s - (  _s\langle\psi|\hat{\rho}_s|\psi\rangle_s )^2$. Therefore the variances of the outcomes $X$ and/or $P$, for the simultaneous measurement of $\widehat{X}$   and   $\widehat{P}$ on $\mathcal{H}$, are larger than $\langle\Delta{\hat{\chi}_s}^2\rangle$ and/or $\langle\Delta{\hat{\rho}_s}^2\rangle$ by 1/4. The additive terms of 1/4 arise from the zero-point fluctuations of the auxiliary system. Similar terms appeared in the Arthurs and Kelly derivation of an uncertainty principle for the simultaneous measurement of position and momentum [20].


\begin{figure}[h]
\centering 
\includegraphics[scale=1.05]{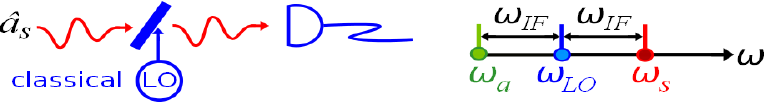}
\caption{Heterodyne Detection }
\end{figure}

In the measurement apparatus of heterodyne detection, the auxiliary noise source is the ``image'' electromagnetic mode (characterized by annihilation operator $\hat{a}_a$) which resides at the same frequency displacement from the classical local oscillator (LO) frequency as does the original mode of interest (characterized by annihilation operator $\hat{a}_s$) as in Fig.\ 2. Even a classical treatment of heterodyning reveals that the beat of the image band with the local oscillator is mapped onto the same detector frequency as the beat of the local oscillator with the original signal. Quantum mechanically, however, we note that we cannot turn the image mode off in the sense that when it is in the vacuum state we have zero-point fluctuations which contribute noise to the detected signal. This noise (essential for preventing the unrealizable perfectly precise  measurement of  $\hat{\chi}_s$ and $\hat{\rho}_s$) is seen to be irrevocably imbedded in the measurement apparatus of the heterodyne detection process.


Extensions of this type are not unique and Naimark's theorem is only meant to recover the fuzzy statistics (not go beyond them). 
Algebraically, we see that for any operator $\widehat{O}_s$ whose commutator with it's adjoint is $\widehat{C}_s \equiv [\widehat{O}_s,{\widehat{O}_s}^\dagger]$ one such extension is $\widehat{E} =  \widehat{O}_s \otimes \widehat{C}_a + {\widehat{C}_s}^\dagger \otimes {\widehat{O}_a}^\dagger$; provided that ${\widehat{C}}^2 = \widehat{C}$ (as it does for $\hat{a}$ and $\hat{A}$). 

Let $\widehat{A}_s$ be the SG operator for the original system of interest; an extension of this form onto $\mathcal{H} = \mathcal{H}_s \otimes \mathcal{H}_a$ is
\begin{equation}
\widehat{Y} \equiv \widehat{A}_s \otimes \widehat{V}_a + \widehat{V}_s \otimes {\widehat{A}_a}^\dagger.
\end{equation}
We can also see explicitly,  from the number representations of $\widehat{A}$ and $\widehat{A}^\dagger$, and recalling the definition of the vacuum projector, $\widehat{V}=|0\rangle\langle0|$, we have
\begin{equation}
\widehat{A} \ \widehat{V} = 0 = \widehat{V} \hspace{.5mm}  \widehat{A}^\dagger
\end{equation}
so that
\begin{equation}
[\widehat{Y}, {\widehat{Y}^\dagger}] = [\widehat{A}_s , {{\widehat{A}_s}^\dagger}] \otimes \widehat{V}_a + \widehat{V}_s \otimes [{\widehat{A}_a^\dagger}, \widehat{A}_a] = 0.
\end{equation}

Next, by solving for the eigenkets of $\widehat{Y}$, we obtain the $\widehat{Y}$ measurement statistics for an arbitrary state, $|\psi \rangle_{s\otimes a} \equiv \Sigma_{n_{s}, n_{a}} \psi_{n_{s}, n_{a}} | n_{s}, n_{a} \rangle$, on $\mathcal{H}$. Setting $\widehat{Y}|Y\rangle=Y|Y\rangle$ yeilds
\begin{equation}
|Y\rangle = \sum_{n_{s} = 0}^{\infty} e^{in_{s} \phi} | n_{s}, 0\rangle + \sum_{n_{a} = 1}^{\infty} e^{-in_{a} \phi} | 0, n_{a}\rangle,
\end{equation}
where $Y=e^{i\phi}$. The eigenvalue, $Y$, could also be zero, in which case the corresponding eigenket can be any superposition of ``off-axis'' number states, i.e., those which do not involve $|n_s, 0\rangle$ or $|0, n_a \rangle$. When the auxiliary mode is in the vacuum state, $|\psi_a\rangle_a=|0\rangle_a$, the general result reduces to the following: the outcome of the $\widehat{Y}$ measurement is a complex number of unit magnitude, $Y=e^{i\phi}$, and the probability distribution for its phase is
\begin{equation}
p(\phi) = \frac{1}{2\pi} |  _s\langle \phi | \psi \rangle_s |^2
\end{equation}
--- identical to that from the POM description of the  measurement associated with $\widehat{A}$.

The single complex-valued outcome, $Y=e^{i\phi}$, of the $\widehat{Y}$ measurement may also be viewed as a pair of real-valued results, $Y_{1}$ and $Y_{2}$, where $Y=Y_1 + iY_2$. The fact that $\widehat{Y}$ commutes with its adjoint implies that its real and imaginary parts, $\widehat{Y}_1\equiv(\widehat{Y}+\widehat{Y}^\dagger )/2$ and $\widehat{Y}_2 \equiv (\widehat{Y} - \widehat{Y}^\dagger)/2i$, are commuting observables. These can also be written in the form $\widehat{Y}_1 = \widehat{C}_s \otimes \widehat{V}_a + \widehat{V}_s \otimes \widehat{C}_a$ and $\widehat{Y}_2 = \widehat{S}_s \otimes \widehat{V}_a - \widehat{V}_s \otimes \widehat{S}_a$ (where $\widehat{C}$ and $\widehat{S}$ are the real and imaginary parts of $\widehat{A}$).  The expected values of these, from a single $\widehat{Y}$  measurement with the auxiliary mode in the vacuum state, are identical to those of the two distinct measurements of $\widehat{C}$ and $\widehat{S}$, i.e., 
\begin{eqnarray}
\langle \widehat{Y}_1 \rangle_{s\otimes a} & = & 
\ _{s}\langle \psi | \widehat{C}_s | \psi \rangle_s    \ _{a}\langle 0 | \widehat{V}_a | 0 \rangle_a \ +\  |\psi_0|^2 \ _{a} \langle 0 | \widehat{C}_a | 0 \rangle_a \nonumber \\ 
 & = & \ _{s}\langle \psi | \widehat{C}_s | \psi \rangle_s
\end{eqnarray}
and similarly
\begin{equation}
\langle \widehat{Y}_2 \rangle_{s\otimes a}
= \ _{s}\langle \psi | \widehat{S}_s | \psi \rangle_s .
\end{equation}
%
%
For the second moments, first note that
\begin{equation}
{\widehat{Y}_1}^2 = {\widehat{C}_s}^2 \otimes \widehat{V}_a + \widehat{V}_s \otimes {\widehat{C}_a}^2 + \frac{{{\widehat{A}_s}^\dagger}\widehat{V}_s}{2} \otimes \frac{\widehat{V}_a \widehat{A}_a}{2} + \frac{\widehat{V}_s \widehat{A}_s}{2} \otimes \frac{{{\widehat{A}_a}^\dagger}\widehat{V}_a}{2}
\end{equation}
and
\begin{equation}
{\widehat{Y}_2}^2 = {\widehat{S}_s}^2 \otimes \widehat{V}_a + \widehat{V}_s \otimes {\widehat{S}_a}^2 - \frac{{{\widehat{A}_s}^\dagger}\widehat{V}_s}{2} \otimes \frac{\widehat{V}_a \widehat{A}_a}{2} - \frac{\widehat{V}_s \widehat{A}_s}{2} \otimes \frac{{{\widehat{A}_a}^\dagger}\widehat{V}_a}{2}.
\end{equation}
Furthermore, since $\widehat{V}\widehat{A}=|0\rangle\langle1|$ and ${\widehat{A}}^{\dagger}\widehat{V}=|1\rangle\langle0|$, the second moments of $\widehat{Y}_1$ and $\widehat{Y}_2$ in a $\widehat{Y}$  measurement with the auxiliary mode in the vacuum state are 
\begin{eqnarray}
\langle \widehat{Y}_1 ^2 \rangle_{s\otimes a} & = &  
\ _s\langle \psi | {\widehat{C}_s}^2|\psi \rangle_s+ |\psi_{0}|^{2} \ _a\langle 0|{\widehat{C}_a}^{2}|0\rangle_a \nonumber \\
 &  = & \ _s\langle \psi |{\widehat{C}_s}^{2}|\psi\rangle_s + \frac{|\psi_{0}|^2}{4}
\end{eqnarray}
and
\begin{equation}
\langle \widehat{Y}_2 ^2 \rangle_{s\otimes a} = \ _s\langle\psi|{\widehat{S}_s}^2|\psi\rangle_s + \frac{|\psi_{0}|^2}{4}
\end{equation}

To interpret the real and imaginary components let us return momentarily to the original notation of the SG operator $\widehat{A} \rightarrow \widehat{e^{i\phi}}$.
The fact that the SG operator does not commute with its adjoint implies that its Hermitian real and imaginary parts, $\widehat{C} \equiv (\widehat{e^{i\phi}} + \widehat{e^{-i\phi}})/2$ and $\widehat{S} \equiv (\widehat{e^{i\phi}} - \widehat{e^{-i\phi}})/2i$ , do not commute: $[\widehat{C}, \widehat{S}] = i\widehat{V}/2$.
Classically we have the trigonometric identity: 
\begin{eqnarray}
& 1 = (e^{i\phi}) (e^{-i\phi}) & = (cos \ \phi + i \ sin \ \phi)(cos \ \phi - i \ sin \ \phi) \nonumber \\
& & = cos^2 \phi + sin^2 \phi.
\end{eqnarray}
Quantum mechanically, we have  $\hat{I} = (\widehat{e^{i\phi}})(\widehat{e^{-i\phi}})$, so 
\begin{eqnarray}
&1 & =  \langle (\widehat{C} + i \widehat{S}) (\widehat{C} - i\widehat{S}) \rangle 
 = \langle \widehat{C}^2 \rangle + \langle \widehat{S}^2 \rangle - i \langle[\widehat{C}, \widehat{S}]\rangle \nonumber  \\
&  &= \langle \widehat{C}^2 \rangle + \langle \widehat{S}^2 \rangle + \frac{|\psi_o|^2}{2}.
\end{eqnarray}
The moments of $\widehat{A}$, or equivalently those of $\widehat{C}$ and $\widehat{S}$ do not correspond to a realizable quantum measurement; but the moments of $\widehat{Y}_{1}$ and  $\widehat{Y}_{2}$ do and from (25), (26) and (29) we obtain

\begin{equation}
\langle{\widehat{Y}_1}^{2}\rangle_{s\otimes a} + \langle{\widehat{Y}_2}^{2} \rangle_{s\otimes a} = 1
\end{equation}
---  in accordance with the assertion that the  outcome of the  $\widehat{Y}$ measurement is a complex number of unity magnitude: $Y=e^{i\phi}$. 

The physical apparatus which realizes the $\widehat{Y}$  measurement has not yet been identified. The presence of the auxiliary mode, however, $must$ (on physical grounds) be an inextricable part of this apparatus --- as it is in heterodyne detection.

\section{Beyond Naimark --- Complementary Phase at the second layer of understanding}
In this section we provide an example of how one can extend the SG operator onto a subset of the two-mode space in order to 
describe a sharp 
quantum phase measurement which yields the next layer of understanding complementarity (be it for phase, time or an angle). There are an infinite number of such subsets  that can be defined on a two-mode space, but a general theory (and final layer of understanding) on an unrestricted  two-mode space will be discussed in section IV.

Historically, $\widehat{Y}$ was derived on physical (rather than algebraic) grounds to achieve a sharp measurement, rather than just recover the fuzzy statistics [7], [8] as follows.
%
Clearly the problems in formulating a time or angle operator stem from the bounded eigenspectra of the complementary quantity (energy or angular momentum). E.g., for a phase operator it's the absence of negative energy states for the quantum harmonic oscillator which leads to the SG operator not commuting with its adjoint. 
The SG operator cannot lower below the vacuum (2) since there are no ``negative-number'' (negative energy) states for the oscillator. The extension, $\widehat{Y}$, however: lowers the original system mode photon number
\begin{equation}
\widehat{Y}|n_{s}\rangle_{s}|0\rangle_{a} = |n_{s} - 1\rangle_{s}|0\rangle_{a} \ \ (n_{s} \ge 1);
\end{equation}
then {\em{continues}} through the vacuum 
\begin{equation}
\widehat{Y}|0\rangle_{s}|0\rangle_{a} = |0\rangle_{s}|1\rangle_{a}
\end{equation}
and raises the auxiliary mode photon number
\begin{equation}
\widehat{Y}|0\rangle_{s}|n_{a}\rangle_{a} = |0\rangle_{s}|n_{a}+1\rangle_{a}.
\end{equation}

Topologically, it is as if $\widehat{Y}$ continues to lower below the vacuum into the auxiliary (negative-number) mode. The visualization of this behavioral aspect can be facilitated by simply relabeling the  number states according to the value of $m=n_{s} - n_{a}$.
One might anticipate that these translations in energy difference will lead to complementarity between $m$ and the ``relative phase'' between the two oscillators.  
Indeed, let us relabel the $\widehat{Y}$ eigenkets as
\begin{equation}
|\phi \rangle ' = \sum_{n_{s}=0}^{\infty} e^{in_{s}\phi} \ |n_{s}\rangle_{s}|0\rangle_{a} +\sum_{n_{a}=1}^{\infty} e^{-in_{a}\phi}| \ 0\rangle_{s}|n_{a}\rangle_{a}.
\end{equation}
These reside on a subset, $\mathcal{H}'$, of $\mathcal{H}_{s} \otimes \mathcal{H}_{a}$ which is spanned by ${|n_{s}\rangle_{s}|n_{a}\rangle_{a} : n_{s}n_{a} = 0}$. 
When the auxiliary mode is in the vacuum state ($n_a = 0$), the $\widehat{Y}$ measurement yields the single-mode statistics and their attendant Paley-Wiener restriction.

We can go {\it{beyond}} these fuzzy statistics by exciting the auxiliary mode to create an arbitrary state on $\mathcal{H}'$:
\begin{equation}
|\psi\rangle = \sum_{n_{s}=0}^{\infty}\psi_{n_{s},0} \ |n_{s}\rangle_{s}|0\rangle_{a} +\sum_{n_{a}=1}^{\infty} \psi_{0,n_{a}} \ |0\rangle_{s}|n_{a}\rangle_{a}.
\end{equation}
Let $\psi_m \equiv \psi_{m,0} \ (\forall  m \ge 0)$ and $\psi_m \equiv \psi_{0,-m} \ (\forall  m < 0)$. The generalized phase wavefunction,
\begin{equation}
\psi' (\phi) \equiv \ '\langle\phi|\psi\rangle = \sum_{m=-\infty}^{\infty} \psi_{m}e^{-im\phi}
\end{equation}
is a two-sided Fourier series. The Paley-Wiener restriction is removed and the $\psi^\prime(\phi)$ can now ``collapse'' to a delta-function. 


Commensurate with its negative-number behavioral aspect, the auxiliary mode can be interpreted as a phase-reversed mode in the following sense. Consider the case of when the auxiliary mode is in the vacuum state $(n_a = 0)$ and denote an initial state by $|\psi \rangle_{0}$. The state (in the Schrodinger picture) after time evolution of an amount $\tau$ is
\begin{equation}
|\psi_{\tau}\rangle = e^{-i(\hat{n}_{s}+\hat{n}_{a})\omega\tau}|\psi_{0}\rangle \big{|}_{n_{a} = 0} = e^{-i\hat{n}_{s} \omega\tau}|\psi_{0}\rangle
\end{equation}
so that the relation between the generalized phase representations of the initial and delayed states is 
simply
\begin{equation}
\psi'_{\tau}(\phi) = \psi'_{0}(\phi + \omega \tau) \ \ (n_{a} =0).
\end{equation}
Now consider the case of the original system being in the vacuum state $(n_s = 0)$. The Schrodinger picture of the delayed version of an initial state $|\psi \rangle_{0}$ is
\begin{equation}
|\psi_{\tau}\rangle = e^{-i(\hat{n}_{s}+\hat{n}_{a})\omega\tau}|\psi_{0}\rangle \big{|}_{n_{s} = 0} = e^{-i\hat{n}_{a} \omega\tau}|\psi_{0}\rangle.
\end{equation}
The initial and delayed generalized phase representations for this case are related by
\begin{equation}
\psi'_{\tau}(\phi) = \psi'_{0}(\phi - \omega \tau) \ \ (n_{s} =0).
\end{equation}
Thus the two modes are phase-reversed in that, under time evolution, the $n_{a}\ge1$ portion of the generalized phase wavefunction moves backwards with respect to the $n_{s}\ge1$ portion.

Of course there is nothing mysterious about the fact that the energy difference, $m$, can be negative. Nor does the phase reversed aspect of the auxiliary mode imply a violation of temporal causality since $\phi$ (which is complementary to $m$) turns out to be the relative phase between the two modes, as we now demonstrate; which will also start to define what we mean by ``relative phase'' which of course implies a quantum measurement.  In so doing we shall also set the stage for the general two-mode relative phase representation (which does not require restriction of the state to $\mathcal{H}'$).


Generalizing complementarity to be viewed as Fourier relations among wavefunctions, we anticipate that what might be of use here is to start with the two dimensional Fourier transform of the $\{\psi_{n_{s},n_{a}}\}$. Indeed, $\Psi(\phi_{s},\phi_{a}) \equiv \ _{s}\langle\phi_{s}| \  _{a}\langle\phi_{a}|\psi\rangle$ provides this:
\begin{equation}
\Psi(\phi_{s},\phi_{a}) = \sum_{n_{s}=0}^{\infty} \sum_{n_{a}=0}^{\infty} \psi_{n_{s},n_{a}} e^{-in_{s}\phi_{s}} e^{-in_{a}\phi_{a}},
\end{equation}
where $| \psi \rangle$ is now an arbitrary state on $\mathcal{H}_{s} \otimes \mathcal{H}_{a}$ with number-ket expansion coefficients  $\psi_{n_{s},n_{a}} \hspace{-.5mm}  \equiv \hspace{-.5mm}  \ _{s}\langle n_{s}|  \ _{a}\langle n_{a}|\psi\rangle$ and $|\Psi (\phi_{s},\phi_{a})|^{2}/(2\pi)^2$ is the probability density function for the simultaneous measurement of $\phi_s$ and $\phi_a$.

Under the change of variables 
\begin{equation}
{\Phi_{\Sigma}} \equiv (\phi_{s} + \phi_{a})/2, \ \ {\Phi_{\Delta}} \equiv (\phi_{s} - \phi_{a})/2, 
\end{equation}
we map to a different wavefunction
\begin{equation}
\psi (\Phi_{\Sigma},\Phi_{\Delta}) = \sum_{j=0}^{\infty} \sum_{m=-j}^{j} \psi_{n_{s},n_{a}} e^{-i(j\Phi_{\Sigma})}e^{-i(m\Phi_{\Delta})}
\end{equation}
where $j \equiv n_{s} + n_{a}$; so that $n_{s} \rightarrow (j+m)/2$ and $n_{a} \rightarrow (j-m)/2$ in the above. Notice that by making the change of variables in the wavefunction (rather than in the PDF) we have also changed the quantum measurement. If instead we made a similar change of variables in the PDF it would correspond to measuring both  $\phi_s$ and $\phi_a$ first and then adding and/or subtracting the results; but that is not what  $\Phi_\Sigma$ and $\Phi_\Delta$ represent in (43).
Since $j$ is bounded from below, the sum phase, $\Phi_{\Sigma}$, is not measureable in the sharp sense. 
In section IV two
reasonable ways of dealing with $\Phi_{\Sigma}$ and hence defining a direct measurement of the relative phase $\phi_{\Delta}$ on $\mathcal{H}_s \otimes \mathcal{H}_a$ are presented. 
For states restricted to $\mathcal{H}'$, one of these ways can be obtained from (43) and we find
\begin{equation}
\psi(\Phi_{\Sigma} =0, \Phi_{\Delta}=\phi)|_{on\mathcal{H}'} = \psi'(\phi)
\end{equation}
which demonstrates that the argument in $\psi^\prime(\phi)$ is a relative phase as asserted. 
We can define a relative phase measurement between any two modes we wish. If however we choose two modes which are already time or phase reversed in some physical sense 
(e.g., an electromagnetic mode of wavevector $\bar{k}$ and another of wavevector $- \bar{k}$) 
then the formalism can lead to more physical insight. For example, if the two modes are the right and left circular polarizations of an electromagnetic plane wave then the relative phase measurement is equal to (not just isomorphic to) the quantum angle measurement.  

 As clarified in the next section, for such photons our definitions of $j$ and $m$ and the change of variables in (42) are appropriate. If instead of such photons our harmonic oscillators (modeling angular momenta) are Schwinger's [21] fermionic primitives then the factors of one-half should go elsewhere (we would divide $j$ and $m$ by two  in the above definitions and multiply by two in (42) for the appropriate change of variables).

\section{Harmonic Oscillator Models of Angular Momenta}

In 1952 Schwinger [21] demonstrated a connection between the algebra of two uncoupled harmonic oscillators and the algebra of angular momenta. In quantum optics this connection has proved useful in the analysis of optical beam splitters [22] although a beam splitter does not actually perform a rotation in physical space and the connection is merely within the mathematics. This connection has also proved useful in calculating the effects of actual rotations on systems but the oscillators (which behave like spin-1/2 bosons) are deemed unphysical [23].   We  put more  physics into this connection by considering rotations of the electromagnetic field. This leads to a subtle but surprisingly  significant modification of Schwinger's model. Also, by describing the oscillator states in the phase representation we will be led to insights on the angles themselves (rather than their conjugate momenta). We begin with a brief summary of the key points of Schwinger's model. Let $\hat{a}_{u}$  and $\hat{a}_{d}$ denote the annihilation operators for two harmonic oscillators which are uncoupled (i.e., independent) so that
\begin{equation}
[\hat{a}_{u}, {\hat{a}_{d}}^{\dagger}] = 0= [\hat{a}_{u}, \hat{a}_{d}].
\end{equation}
Defining
\begin{equation}
\hat{J}_{+} \equiv {\hbar} {\hat{a}_{u}}^{\dagger}{\hat{a}_{d}}, \hspace{2mm} \hat{J}_{-} \equiv {\hbar} {\hat{a}_{d}}^{\dagger}{\hat{a}_{u}} \ \ \text{and} \ \ \hat{J}_{z} \equiv \frac{\hbar}{2}(\hat{\eta}_{u}-\hat{\eta}_{d}),
\end{equation}
it is easy to show that
\begin{equation}
[\hat{J}_{+}, \hat{J}_{-}] = 2{\hbar} \hat{J}_{z} \ \ \text{and} \ \ [\hat{J}_{+}, \hat{J}_{-}] = \pm \hbar \hat{J}_{\pm},
\end{equation}
which are the fundamental communication relations of angular momentum. The raising and lowering operators, $\hat{J}_{\pm}$, can of course be alternatively expressed in terms of the x and y component angular momentum operators as
\begin{equation}
\hat{J}_{\pm}=\hat{J}_{x} \pm i \hat{J}_{y}.
\end{equation}

Since, from (43), $\hat{J}_{\pm}$ raise and lower the eigenvalue of $\hat{J}_{z}/\hbar$ (i.e.,  $m$) by one, it is as though we have one spin-1/2 particle, with spin up (or down) associated with each quanta of the u (or d) oscillators. The z component of angular momentum is then simply $\hbar/2$ times the difference in the number of up and down quanta, commensurate with (43). Therefore, in terms of eigenvalues we have
\begin{equation}
m=(n_{u}-n_{d})/2 ,
\end{equation}
from which we anticipate $j = (n_{u}+n_{d})/2$, where $\hbar^{2} j (j+1)$ is the eigenvalue of $\hat{J}^{2} \equiv {\hat{J}_{x}}^{2} +{\hat{J}_{y}}^{2}+{\hat{J}_{z}}^{2}$. Indeed, 
\begin{equation}
\hat{J}^{2} = \frac{1}{2}(\hat{J}_{+}\hat{J}_{-} +\hat{J}_{-}\hat{J}_{+}) +{\hat{J}_{z}}^{2} = {\hbar}^{2} (\frac{\hat{n}_{u}+\hat{n}_{d}}{2})(\frac{\hat{n}_{u}+\hat{n}_{d}}{2} +1)
\end{equation}
so that
$
j = (n_{u}+n_{d})/2
$
as expected.

In Schwinger's scheme however, a total of two up and down quanta will always yield $ j = 1$, i.e., the anti-symmetric singlet state (of $j=0$) would never occur.   As Sakurai 
[23] 
puts it: ``only totally symmetrical states are constructed by this method. The primitive spin 1/2 particles appearing here are actually {\it{bosons}}! This method is quite adequate if our purpose is to examine the properties under rotations of states characterized by $j$ and $m$ without asking how such states are built up initially.'' 
Since spin-1/2 particles must obey Fermi statistics we shall not attempt to make physical sense of these {\it{spin-1/2 primitives which  act like bosons}} and shall instead simply refer to them as fermionic primitives (mathematical entities which need not be represented in the physical world).

The photon is a boson which nonetheless resembles a fermion in the sense that its spin space is two dimensional, i.e., it is ``spin-1 with m = 0 missing'' [24]. Indeed, two harmonic oscillator modes are sufficient to describe the polarization state of a single k-vector component of an electromagnetic wave since we need only consider the transverse components of its vector potential. Therefore, It seems reasonable to attempt to reconstruct the algebra of angular momenta from these physically significant photonic primitives. We pursue this  by considering rotations of the electromagnetic field.

Let $\hat{\bar{A}}$ be the vector potential operator for an electromagnetic wave comprised of the two circularly polarized, $\bar{z}$ propagating, same frequency modes. 
By requiring that the expected value of this vector operator transform like a classical vector under rotations, we obtain the
 well-known [24] results:
\begin{equation}
\widehat{R}_{z}(\phi) \hspace{.5mm} {\hat{a}_{r}}^{\dagger} \hspace{.5mm} {\widehat{R}_{z}}^{\dagger}(\phi) = {\hat{a}_{r}}^{\dagger}e^{-i\phi}
\hspace{1mm}
\text{and}
\hspace{1mm}
\widehat{R}_{z}(\phi) \hspace{.5mm} {\hat{a}_{l}}^{\dagger} \hspace{.5mm} {\widehat{R}_{z}}^{\dagger}(\phi) = {\hat{a}_{l}}^{\dagger}e^{i\phi}.
\end{equation}
From this Heisenberg picture of a rotation about the z-axis by an amount $\phi$ we have, in the Schrodinger picture, that
\begin{equation}
e^{-i\hat{J}_{z}\phi/ \hbar}|1\rangle_{r} = e^{-i\phi}|1\rangle_{r},
\end{equation}
i.e., a right handed circularly polarized photon is an eigenstate of $\hat{J}_{z}/\hbar$ with eigenvalue $m= +1$, where we used the assumption that the vacuum is rotationally invariant, ${\hat{R}_{z}}^{\dagger}(\phi)|0\rangle = |0\rangle$, and $|1\rangle_{r}$ is $|1,0\rangle \equiv {\hat{a}_{r}}^{\dagger}|0,0\rangle$ in $|n_{r},n_{l}\rangle$ notation. Similarly, a left handed circularly polarized photon is associated with $m = -1$ and therefore photons are said to be particles of spin 1 with $m = 0$ missing. Furthermore, from (51) we find that the Schrodinger picture of a rotation about the z-axis for an arbitrarily polarized field (expressed in the circularly polarized basis) is 
\begin{equation}
\widehat{R}_{z} (\phi)|\psi \rangle = \sum_{n_{r}, n_{l}} \psi_{n_{r}, n_{l}}e^{-i(n_{r} - n_{l})\phi}|n_{r}, n_{l}\rangle
\end{equation}
which we notice is physically indistinguishable from a differential phase shift of the two circularly polarized modes. Therefore, when we utilize this rotation to derive its complementary angle-kets, as in Appendix 2; for this choice of modes we would also obtain the relative phase-kets (similarly deriveable under differential phase shift).   {\it{Thus, for this particular choice of mode set, the angle and phase measurements are exactly identical (rather than merely isomorphic).}} In any event, the connection between angular momentum and these photonic primitives is clearly 
\begin{equation}
\hat{J}_{z} = \hbar (\hat{n}_{r} - \hat{n}_{l}).
\end{equation}
Note the absence of the factor of 1/2 which was present in the case of fermionic primitives.  As (54) leads to $m=n_{r} - n_{l}$ we expect $j=n_{r}+ n_{l}$. However because we've scaled up $\hat{J}_{z}$ (by 2) we find that we must also scale up $\hat{J}_{x}$ and $\hat{J}_{y}$ (by 2) in order to make sense of $\hat{J}^{2}$. We can introduce the scaling either in the relation between $\hat{J}_{x,y}$ and $\hat{J}_{+,-}$; or in the definitions of $\hat{J}_{+}$ and $\hat{J}_{-}$. Choosing the latter:
\begin{equation}
\hat{J}_{+} \equiv 2\hbar \ {\hat{a}_{r}}^{\dagger} \ \hat{a}_{l} \ \ \text{and} \ \ \hat{J}_{-} \equiv 2\hbar \ {\hat{a}_{l}}^{\dagger} \ \hat{a}_{r}
\end{equation}
 and (48) still holds. This leads to 
\begin{equation}
\hat{J}^2 = {\hbar}^2 (\hat{n}_r + \hat{n}_l)(\hat{n}_r + \hat{n}_l + 1) 
\end{equation}
so that $j = n_r + n_l$ as desired. An unavoidable consequence of this scaling is the appearance of a factor of 2 in the commutators
\begin{equation}
[\hat{J}_{i},\hat{J}_{j}]= 2i\hbar \ \epsilon_{ijk} \ \hat{J}_{k} \ \ (i,j,k \ \epsilon \ \{ x,y,z \}).
\end{equation}
This however is still the same group (although the structure constant has doubled) i.e., we {\it{have}} reproduced the algebra of angular momenta with these physically interpretable photonic primitives. What has happened is perhaps more clear when viewed in terms of the commutators involving $\hat{J}_{\pm}$.

We now have
\begin{equation}
[\hat{J}_{+},\hat{J}_{-}]= 4\hbar \hat{J}_{z},
\end{equation}
where the 4 comes from scaling up the $\hat{J}_{\pm}$, but there is no way to scale these to alter the fact that we now have
\begin{equation}
[\hat{J}_{z},\hat{J}_{\pm}]= \pm 2\hbar \hat{J}_{\pm}.
\end{equation}
The factor of two that has to appear in (59) is a solution rather than a problem however since it means that $\hat{J}_{\pm}$ will raise and lower the eigenvalue of $\hat{J}_{z}/\hbar$ by two rather than one --- which is exactly what we want! We can also see this from (55) which indicates that $\hat{J}_{-}$ (for example) will annihilate one right handed photon (with z-component angular momentum of $\hbar$) and create one left handed photon (with z-component angular momentum of $-\hbar$) thereby lowering the value of $m$ by two. Thus $\hat{J}_{-}$ lowers the $m=+1$ state immediately to the $m=-1$ state while automatically skipping over the $m=0$ case which does not exist for a photon.

In Fig.\ 3 we indicate the allowed photonic states by points (solid circles) in the $n_{r}, n_{l}$ plane; which is also labeled by $j$ and $m$. Notice that for an odd (or even) total number of photons $m$ must also be odd (or even). The so-called missing states are indicated by empty circles. For ordinary bosons we would ``fill in these holes,'' and as we can still use the angle representation to describe their polarization state we maintain the connection with harmonic oscillators, but then those would have to be the presumed unphysical fermionic primitives for which we would use the $n_{u}/2$ and $n_{d}/2$ axes.

Consider for example the quantum angle representation of the state of a single particle (labeled ``particle'' in Fig.\ 3). Formally, the angle of a particle requires a field for its measurement (else the finite dimensional state space would not permit wavefunction collapse [25]). When the field (also labeled in Fig.\ 3) is comprised of photons (rather than up/down oscillators) the state space for the complete, sharp, description of this measurement is a composite of the state space of the particle and that of the field (for which we could also include the vacuum state in order to turn the field ``off''). For this field/particle system to go beyond the fuzzy statistics the state of the field would have to be entangled with that of the particle prior to the measurement. Otherwise (e.g., if the field was ``off'' prior to the measurement) the outcomes will be Paley-Weiner limited and these ``single-particle'' fuzzy statistics can be obtained via a simple Fourier transform: $\psi(\varphi)=\sum_{m=-j}^{j} \psi_{j,m} \  e^{- i m \varphi}$, where $\varphi$ is the angle about the z-axis.
The fact that the quantum angle distribution cannot vanish over an interval of non-zero width has some interesting physical consequences 
such as \textit{spin up really does point up}
 in [26]. 
These phenomena and the necessity of a field etc., would not be revealed if we restricted our attention to a different measurement --- the discrete-angle measurement [27] which can be described by an Hermitian operator on the space of a single particle via a wrap-around term (depicted by the semi-circle in Fig. 3) akin to that of Pegg and Barnett [9], [10]. Unlike the case of discrete-phase convergence to the fuzzy single-mode phase statistics however, the discrete-angle statistics cannot converge to the fuzzy single-particle angle statistics since we can't just take the limit $j \rightarrow \infty$ and still be referring to a spin-$j$ particle. Moreover the discrete-angle statistics are complementary to a periodically replicated version of the angular momentum spectra. But real angles are continuous --- just as real angular momenta are not  periodic.

\begin{figure}[h]
\centering 
\includegraphics[scale=.4]{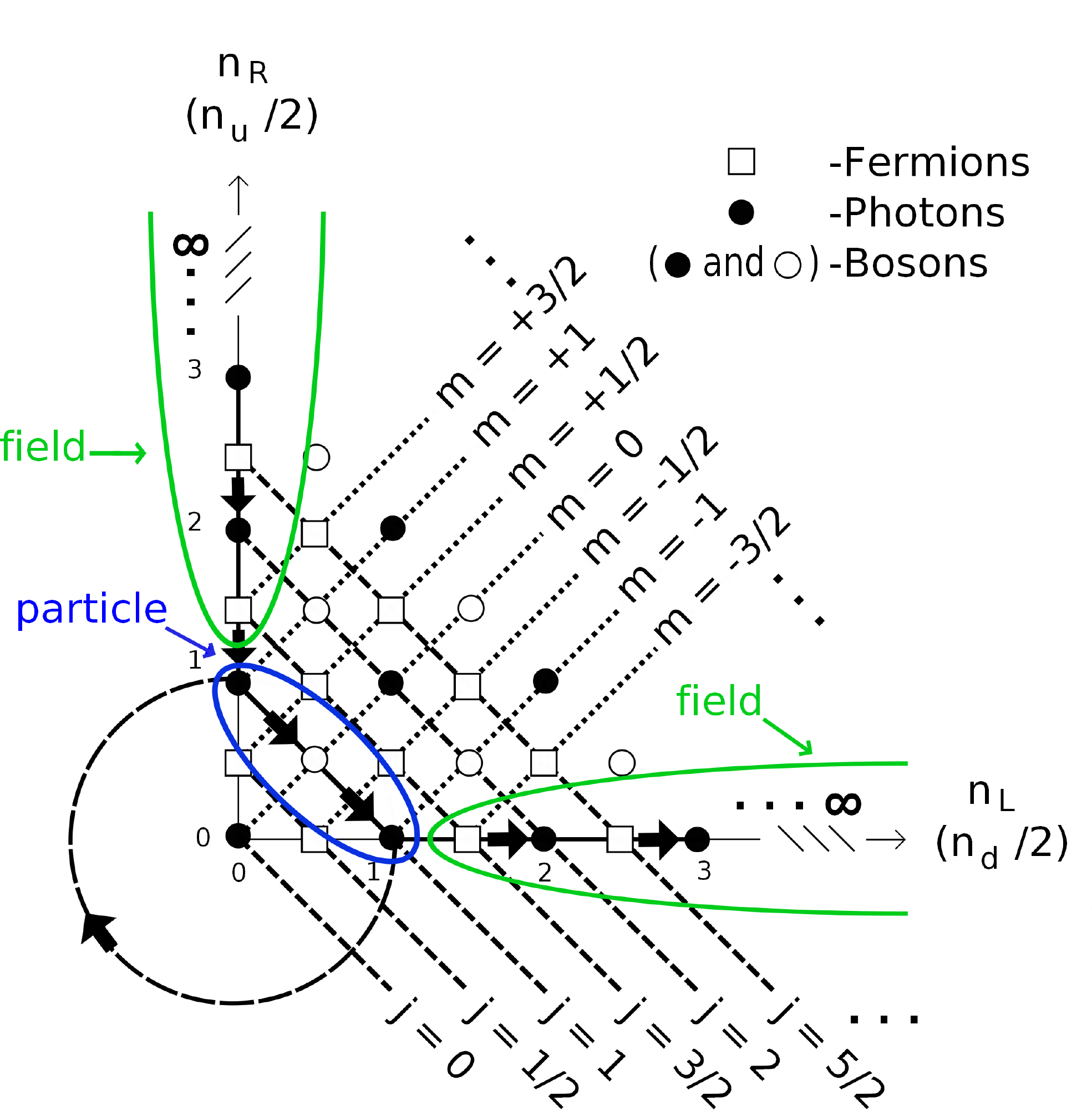}
\caption{Harmonic Oscillator Models of Angular Momentum}
\vspace{-2mm}
\end{figure}

\vspace{-2mm}

\section{a general quantum theory of angle and relative phase measurement}

 There are an infinite number of  subsets such as $\mathcal{H}'$, or the field/particle space in Fig.\ 3, that can be defined on a two-mode space wherein each value of $m$ also corresponds to a single value of $j$. The general theory (and final layer of understanding) on an unrestricted  two-mode space, 
presented herein, can cover all of those infinite number of possible measurements and it demonstrates two reasonable ways of dealing with degeneracy in $j$. 
These also correspond to 
the two
reasonable ways of dealing with the uncollapsible $\Phi_{\Sigma}$ and hence defining a direct measurement of the relative phase $\phi_{\Delta}$ on $\mathcal{H}_s \otimes \mathcal{H}_a$.
\newline
\indent
In probability theory, to eliminate one variable from a two-dimensional PDF there are two reasonable choices: form a conditional PDF; or form a marginal PDF. Rather than doing this on the PDF we extrapolate and apply the concepts to a POM to form either a ``conditional measurement,'' i.e., a snapshot in absolute time of the relative phase; or a ``marginal measurement,'' i.e., a time average of the relative phase distribution.
\newline
\indent In quantum theory, there are two reasonable alternatives: if the final states of a measurement are distinguishable we add probabilities; if the final states of a measurement are indistinguishable we add probability amplitudes [28]. 
It is perhaps initially surprising that these two things reasonable (from two totally different perspectives) coincide. I.e., we will show that taking a snapshot corresponds to adding amplitudes and taking a time average corresponds to adding probabilities. 
\newline
\indent
If states of different $j$ correspond to different (distinguishable) particles then we could impose the ``add probabilities'' constraint and equivalently  argue that the snapshot measurement cannnot be realized for such a system. Otherwise we could argue that as designers of quantum measurements we are free to choose to do a measurement that yields final states which are distinguishable (in $j$) or not. Specifically, let the system be right and left circularly polarized photons so that the oscillators have physical significance and the relative phase measurement is equivalent to the quantum angle measurement. In a snapshot measurement of this angular distribution (say at $\Phi_\Sigma = 0$) how are we to tell which $j$-branch contributed a result? The snapshot measurement has a connection to (43) from which we see that information on $j$ vanishes when we take $\Phi_\Sigma = 0$. It is as if ``we don't take enough time'' to distinguish the different $j$. On the other hand, the time average measurement is a marginal POM (rather than a marginal PDF) so the connection with (43) is not as direct but one can see that when $\Phi_\Sigma$ varies the differences in $j$
can have an effect --- making them distinguishable (even if the entire system is comprised of indistinguishable photons) and so it is palatable that we end up adding probabilities. It is as if ``distinguishability is a matter of how long we look.'' Examples will clarify this after we first present the details of the formalism. 
\newline
\indent
Let 
\vspace{-2mm}
\begin{equation}
|\Phi_{\Sigma},\Phi_{\Delta}\rangle \equiv \sum_{j=0}^{\infty} \sum_{m=-j}^{j} |n_{s},n_{a}\rangle \ e^{i(j\Phi_{\Sigma})} e^{i(m\Phi_{\Delta})}
\end{equation}
where for photons: $n_{s} \rightarrow (j+m)/2$ and $n_{a} \rightarrow (j-m)/2$ in the above and $m$ increments by two in the sum. 
We can eliminate $\Phi_\Sigma$ to obtain a {\it{marginal measurement}} of $\Phi_\Delta$ on $\mathcal{H}_{s} \otimes \mathcal{H}_{a}$ by applying an ``absolute time average'' to $|\Phi_\Delta,\Phi_\Sigma\rangle \langle\Phi_\Delta,\Phi_\Sigma|$, resulting in  the marginal POM:
\begin{align}
& (2\pi) \ d\hat{\Pi}_M (\Phi_\Delta) \equiv \int_{-\pi}^{+\pi} \frac{d\Phi_\Sigma}{2\pi}\ |\Phi_\Delta,\Phi_\Sigma\rangle \langle\Phi_\Delta,\Phi_\Sigma| =  \nonumber \\
& \sum_{j=0}^{\infty} \left[\left(\sum_{m=-j}^{+j} |j,m\rangle \ e^{im\Phi_\Delta}\right)\left(\sum_{m'=-j}^{+j} \langle j, m'| \ e^{-im'\Phi_\Delta}\right)\right]. 
\end{align} 
Because both of the inner sums use the same value of $j$, interference among states of different $j$  is excluded and we have (for pure states) the following probability distribution function:
\begin{equation}
P_M(\Phi_\Delta) = T_r[\hat{\rho} \ d\hat{\Pi}_M(\Phi_\Delta)] = 
\left(\frac{1}{2 \pi}\right) 
\displaystyle\sum\limits_{j=0}^{\infty} | \Psi^{(j)}(\Phi_\Delta)|^2 
\end{equation}
where $T_r$ denotes trace; $\hat{\rho}$ is the density matrix; and
\begin{equation}
\Psi^{(j)}(\Phi_\Delta) \equiv \displaystyle\sum\limits_{m=-j}^{+j} \Psi_{j,m} \ e^{-im \Phi_\Delta} 
\end{equation}
is the quantum angle representation for each $j$-branch of expansion coefficients
$\Psi_{j,m} = \langle j,m|\psi \rangle$. Thus, in this marginal or time averaged measurement of $\Phi_\Delta$, the results from
states of different $j$ are distinguishable and we are led to adding probabilities in (62). 

We also might  eliminate $\Phi_\Sigma$ to obtain a {\it{conditional measurement}} of $\Phi_\Delta$ on $\mathcal{H}_{s} \otimes \mathcal{H}_{a}$ by taking ``snapshot in absolute time'' via conditioning  $|\Phi_\Delta,\Phi_\Sigma\rangle \langle\Phi_\Delta, \Phi_\Sigma|$ to  $\Phi_{\Sigma}=0$ (for example) resulting in  the conditional  POM:
\begin{align}
& (2\pi C) \ d\hat{\Pi}_C(\Phi_\Delta) \equiv 
 \ |\Phi_\Delta,\Phi_\Sigma =0\rangle \langle\Phi_\Delta,\Phi_\Sigma = 0| = \nonumber \\
& \left(\displaystyle\sum\limits_{j=0}^{\infty}\displaystyle\sum\limits_{m=-j}^{+j} |j,m\rangle \  e^{im\Phi_\Delta}\right) 
\left(\displaystyle\sum\limits_{j'=0}^{\infty}\displaystyle\sum\limits_{m'=-j'}^{+j'} \langle j',m'| \ e^{-im'\Phi_\Delta}\right) 
\end{align}
where the renormalization constant $C$ is:
\begin{equation}
C \equiv P (\Phi_\Sigma =0) = \displaystyle\sum\limits_{m=-\infty}^{\infty} \left\{\left|\left(\displaystyle\sum\limits_{j=0}^{\infty} \Psi_{j,m}\right)\right|^2\right\} 
\end{equation}
 the probability of the conditioning event.


In contrast to the case in (61) the sums over $m$ in (64) now use different values of $j$, thereby permitting interference among the states of different $j$. Therefore  we have (for pure states)
 the probability distribution function:

\vspace{-4mm}
\begin{equation}
P_C(\Phi_\Delta) = T_r[\hat{\rho} \ d\hat{\Pi}_C(\Phi_\Delta)] = \left(
\frac{1}{2\pi C}
\right) \left|\displaystyle\sum\limits_{j=0}^{\infty} \Psi^{(j)}(\Phi_\Delta)\right|^2 
\end{equation}
so that in this {\it{conditional}} ``snapshot'' measurement of $\Phi_\Delta$
we are adding amplitudes, the $\Psi^{(j)}(\Phi_\Delta)$,
before taking the magnitude square in (66).

Note that the snapshot measurement recovers the $\widehat{Y}$ measurement on $\mathcal{H}'$, or any other of the infinite number of operators that could be defined on subsets of $\mathcal{H}_s \otimes \mathcal{H}_a$ in which each value of $m$ corresponds to a unique value of $j$ (for these non-degenerate cases $C = 1$ so that renormalization would not be required). 
For the degenerate cases we could equivalently form an amplitude for being in a state of $m$, independent of $j$ via: $\Psi_{m} \equiv \sum_{j} \Psi_{j,m}$ and Fourier transform these to form the wavefunction underlying the snapshot PDF of (66). 
We could similarly define a conditional measurement of $\Phi_\Delta$ by taking a snapshot at some other value of $\Phi_{\Sigma}$ but that would be equivalent to applying the above procedure to a state that has undergone time evolution of that same (conditioning) amount $\Phi_{\Sigma} /\omega = t$ --- which is computationally  easier (where $\omega$ is the radian frequency of our oscillator, which we now set equal to one for simplicity and to emphasize that $\Phi_{\Sigma}$ corresponds to absolute time). 

\section{quantum angle representations of optical polarization}


Applying the general formalism to a case that can naturally support a  time-reversal symmetry (as in section II) should yield more physical insights.
Herein we take the original system (s) and the auxiliary mode (a) to be the right (R) and left (L) circularly polarized modes of a single k-vector
z-propagating
plane wave
so that the 
relative phase measurement  is the quantum angle measurement in a system comprised of physically realizable primitives.
The conditional measurement now corresponds to taking a snapshot of the angular distribution of the electric field about the z-axis 
(at some angle $\phi = \Phi_{\Delta}$ with respect to the x-axis) taken at some time $t = \Phi_{\Sigma}$. The marginal measurement forms a properly weighted time average of these angular snapshots which trace out, and thereby result in, a quantum version of the polarization ellipse.

When the state of the field is comprised of a single value of total angular momentum (i.e., when the only non-zero probability amplitudes have one unique value of $j = {n_R + n_L})$ the polarization ellipse is the same distribution as a snapshot taken at any point in absolute time. Moreover, in the case of an x-polarized number state, with the y-polarized mode in the vacuum state, one would expect these identical distributions to peak at both $\phi=0$ and $\phi=\pm \pi$. For example, one x-polarized photon, $\it{or}$ two x-polarized photons, should have a polarization ellipse along the x-axis with ``up along x'' and ``down along x'' being equally most likely; and the identical snapshot distribution at any time must follow suit. Indeed this physically reasonable result holds in the quantum angle representation. 

However, if we have a superposition of one x-polarized photon $\it{and}$ two x-polarized photons then the value of $j$ is not unique and the snapshot distributions are not identical to the polarization ellipse. For example, the superposition 
\begin{eqnarray}
&\sqrt{2} \ |\psi \rangle =| & 1\rangle_{x}|0\rangle_{y} + |2\rangle_{x}|0\rangle_{y} 
= (|1\rangle_{R}|0\rangle_{L} + |0\rangle_{R}|1\rangle_{L})/\sqrt{2} \nonumber \\ 
& & \hspace{-1mm} +  \ (|2\rangle_{R}|0\rangle_{L} + |0\rangle_{R}|2\rangle_{L} + \sqrt{2} |1\rangle_{R}|1\rangle_{L})/2
\end{eqnarray}
has an angle representation comprised of $j=1$ and $j=2$ components given by $\sqrt{2} \ \psi(\phi) = \Psi^{(1)}(\phi) + \Psi^{(2)}(\phi)$ where
\begin{equation}
 \Psi^{(1)}(\phi) = \sqrt{2} \ \rm{Cos}(\phi) \hspace{2mm} \rm{and} \hspace{2mm} 
\Psi^{(2)}(\phi) =  Cos(2\phi) + 1/\sqrt{2}.
\end{equation}
If we magnitude square either of these it will have a peak at $\phi=\pi$ and the sum of those probabilities would yield the polarization ellipse. If we add amplitudes instead, then although both $j$-components have an amplitude for being ``down along x'' at this time (when $\Phi_\Sigma =0$) those amplitudes are out of phase at this time so they cancel and the snapshot distribution (when $\Phi_\Sigma =0$) is only ``up along x'' (i.e., peaks only at $\phi=0$) as shown in Fig.\ 4, where
the x-axis is absolute time ($\Phi_\Sigma$ from 0 to $\pi$); the y-axis is the angle ($\phi$ from $-\pi$ to $\pi$); and the contours indicate the $\it{angular}$ probability density (from $0.1$ to $0.8$). I.e., each ``slice in x'' is a (normalized) snapshot PDF.  Later (when $\Phi_\Sigma=\pi$) the snapshot distribution must reflect an angular distribution of electric field vectors that is primarily ``down along x'' (i.e., the snapshot distribution, when $\Phi_\Sigma=\pi$, must peak at $\phi=\pm \pi$ --- as it does). 

\begin{figure}[h]
\includegraphics[scale=.65]{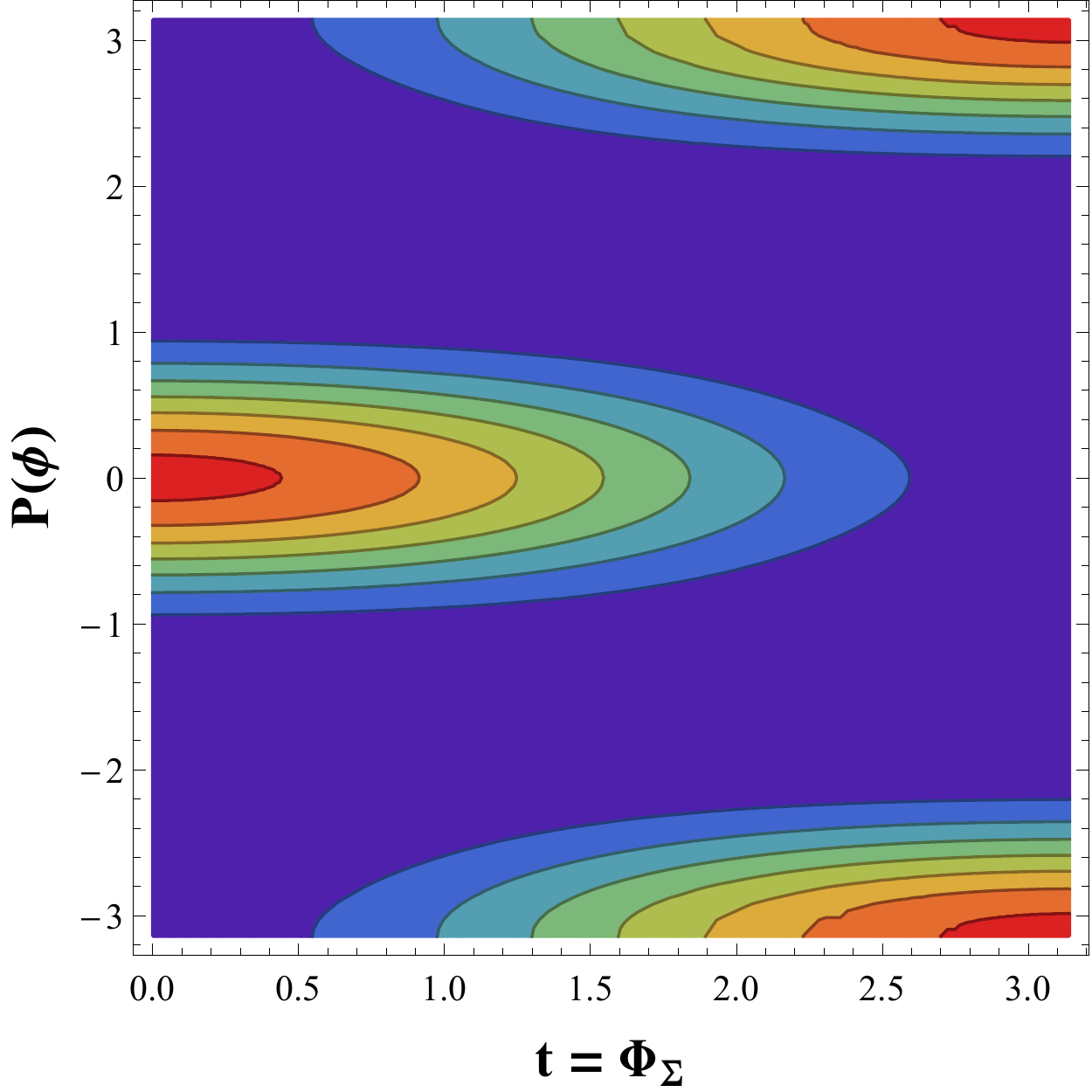}
\caption{A Sequence (at times $t$) of Snapshot PDFs (along y) for a Superposition of One and Two X-Polarized Photons.}
\vspace{-2mm}
\end{figure}

Consider now how ``up along x'' can evolve into ``down along x.'' Classically, the two (right and left handed) circular polarization vectors counter-rotate in time and their sum creates a linearly polarized vector that shrinks to zero half-way between the time it is ``up along x'' and the time it is ``down along x.'' Quantum mechanically, the renormalized snapshot PDFs cannot vanish (although without renormalization these can shrink towards zero, reflecting an unlikelihood of taking a snapshot at that time).
Moreover, these PDFs are Paley-Weiner restricted in shape (for finite $j$) so delta-functions are not allowed and the PDFs cannot equal zero over angular intervals of non-zero length. It is analogous to considering how one might deform a balloon from one pointing up to one pointing down without breaking the balloon, while also maintaining symmetry in $\phi$. Two possibilities are to have a peak at $\phi=0$ come down and spread out into a more uniform shape while the peak at $\phi=\pm \pi$ comes up; or the peak at $\phi=0$ can come down and spread out into two or more discernable counter-rotating peaks, which then recombine into a peak at $\phi=\pm \pi$. The superposition of one and two x-polarized photons of (67) gives an example of the former in Fig.\ 4; and an x-polarized coherent state of $N=1$ gives an example of the later in Fig.\ 5.

\begin{figure}[h]
\centering 
\includegraphics[scale=.65]{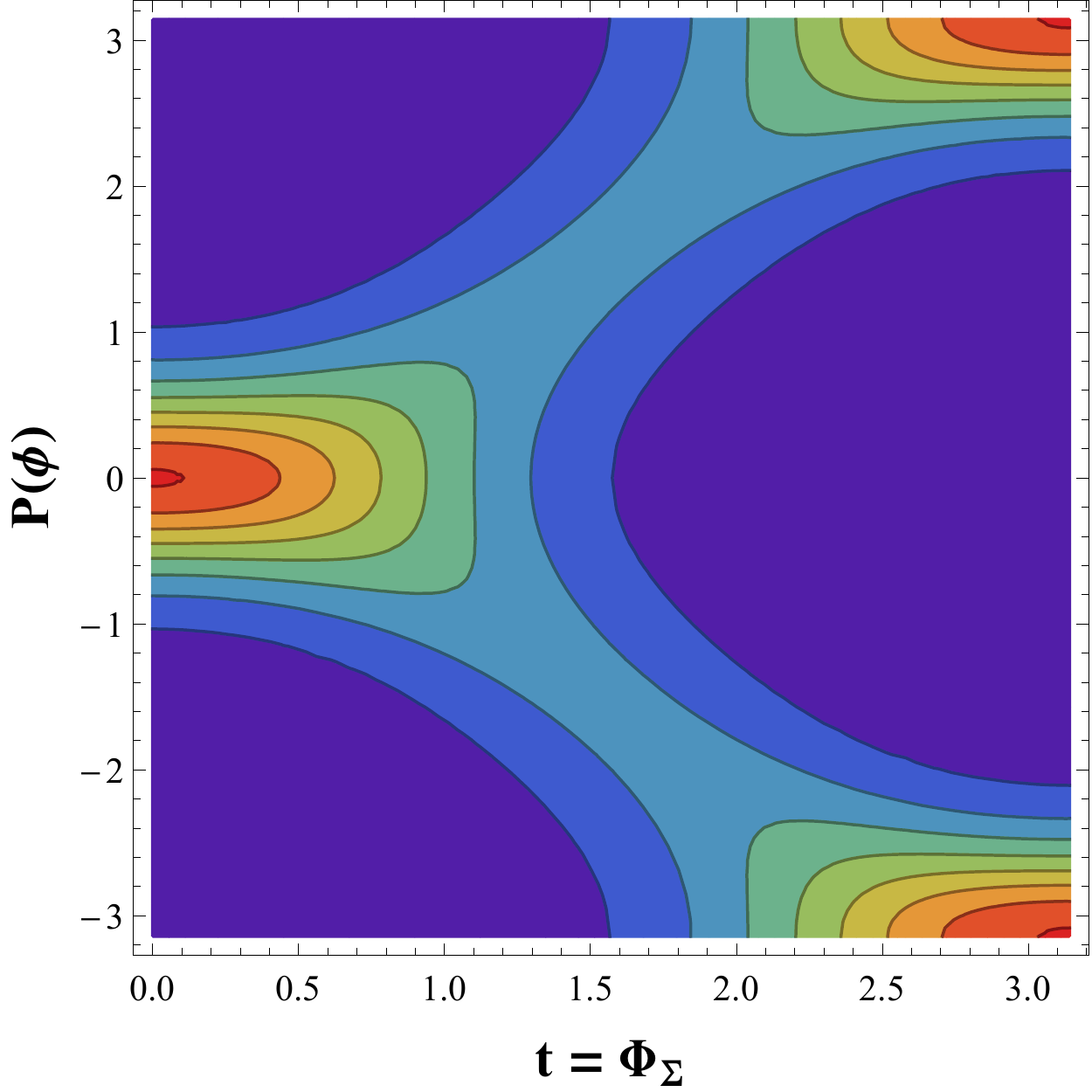}
\caption{A Sequence (at times $t$) of Snapshot PDFs (along y) for a Coherent State of $N = 1$.}
\end{figure}

\begin{figure}
\centering 
\includegraphics[scale=.64]{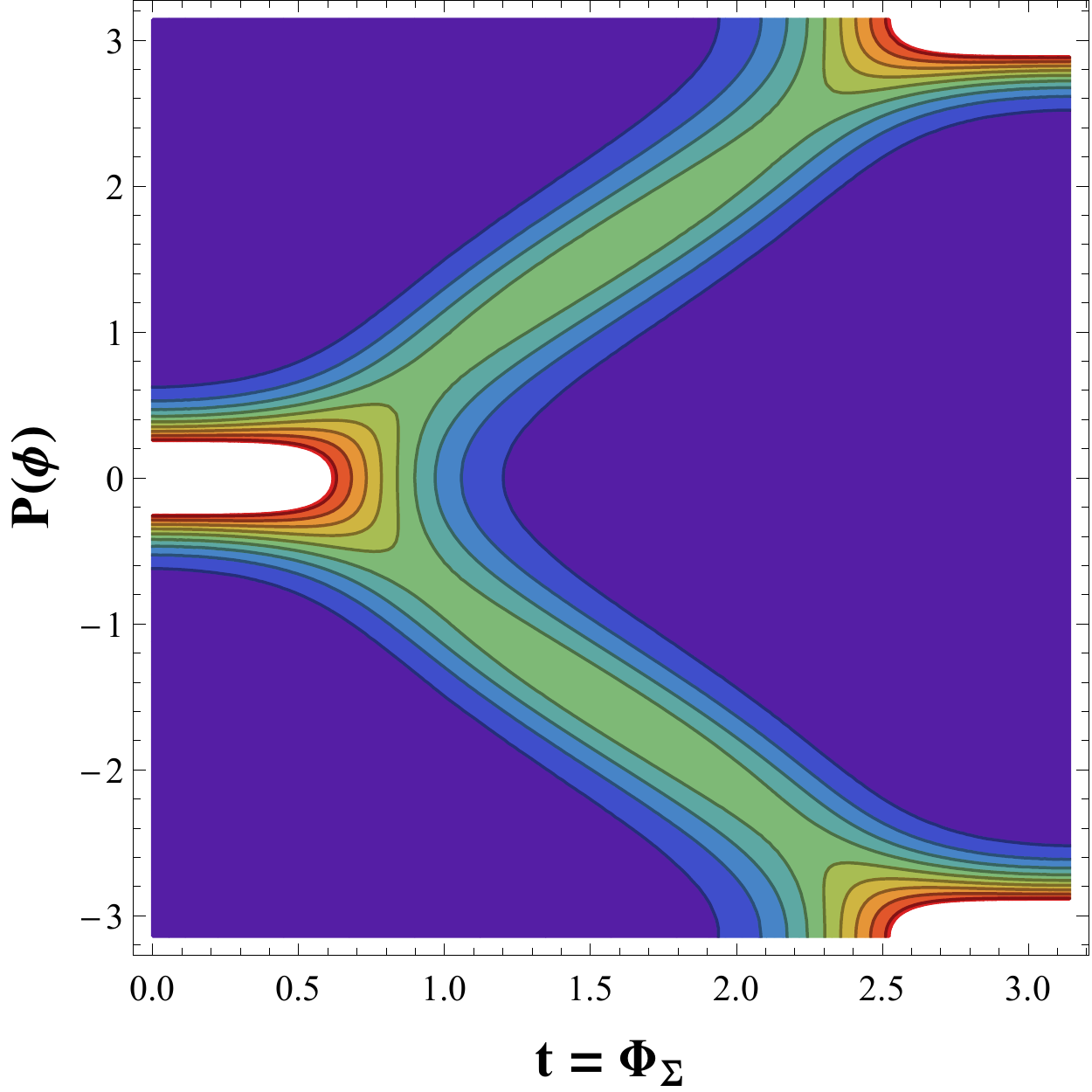}
\caption{A Sequence (at times $t$) of Snapshot PDFs (along y) for a Coherent State of $N = 4$.}
\vspace{-3mm}
\end{figure}

Figures 6 and 7 similarly depict the evolution of the snapshot PDFs for x-polarized coherent states of $N= 4$ and  $N=9$, respectively. Both demonstrate again the splitting of the initial ``up along x'' peak into two discernable peaks (which would be counter-rotating in a polar plot) which then again have to recombine into a single ``down along x'' peak (at $\phi = \pm \pi$) when $\Phi_\Sigma \rightarrow \pi$. For $N=9$, Fig.\ 7 also reveals that when $\Phi_\Sigma$ is near $\pi/2$ two more peaks or side-lobes (at $\phi=0$ and  at $\phi = \pm \pi$) become visible. 



\begin{figure}[h]
\centering 
\includegraphics[scale=.64]{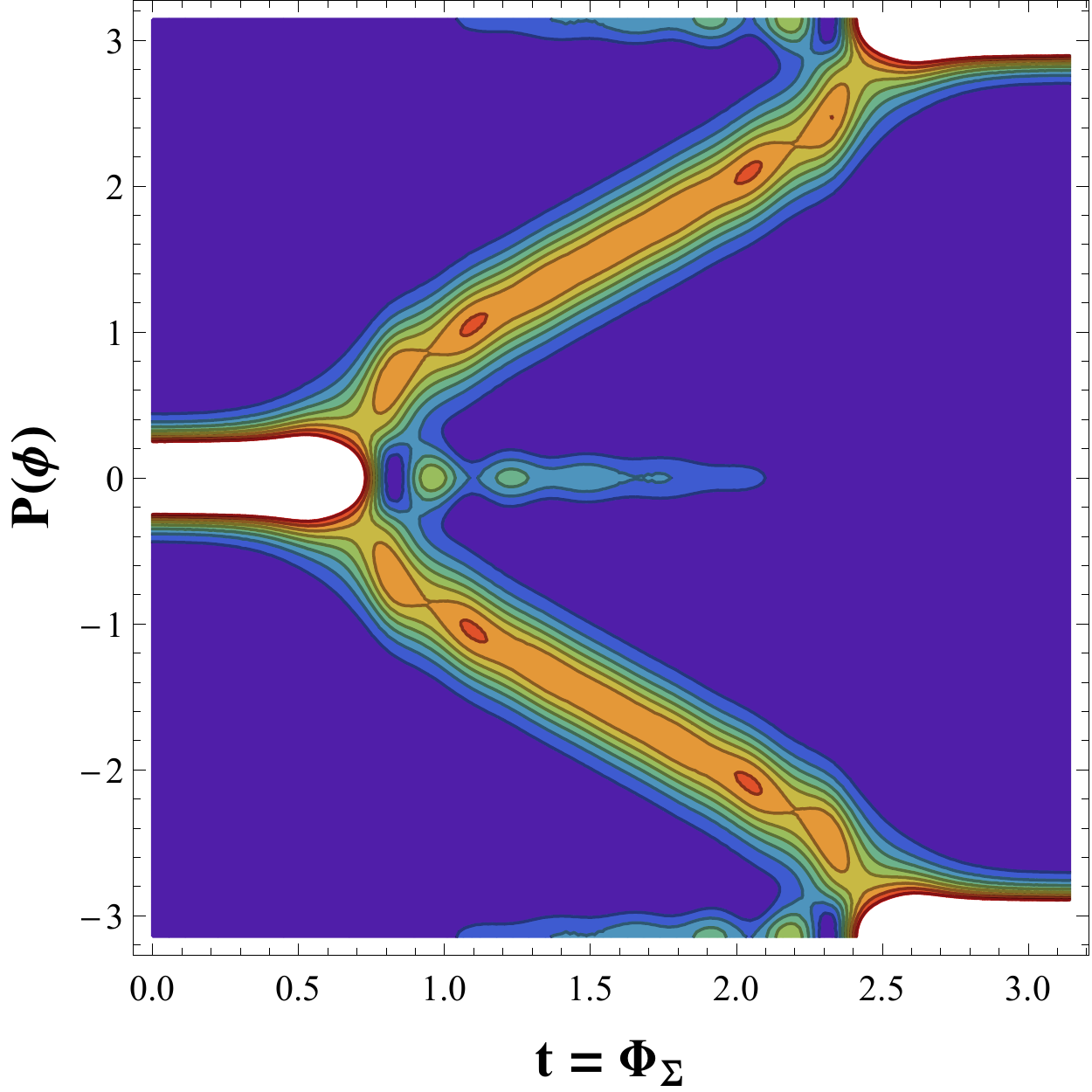}
\caption{A Sequence (at times $t$) of Snapshot PDFs (along y) for a Coherent State of $N = 9$.}
\end{figure}

Before taking up the issue of the behavior at $\phi=\pi/2$ it is useful to consider the role that the probability of $\Phi_\Sigma$ plays in the relation between a sequence of snapshots and the quantum polarization ellipse. 
For example, the snapshot PDF for an x-polarized coherent state of $N=9$ reveals an appreciable $P(\phi=\pi/2)$ when the sum phase  is near $\pi/2$, as shown in Fig.\ 8 which presents
three snapshot PDFs in a polar plot on a linear scale. 
The peak of the snapshot PDF at time $t=\pi/2$ is down from the peak of the snapshot PDF at time $t=0$ by less than a factor of four. However, the probability of taking a snapshot at time $t=\pi/2$ is over three orders of magnitude smaller than the probability of taking a snapshot at time $t=0$. Thus, when we take the time average inherent in forming the polarization ellipse the influence of the $t=\pi/2$ snapshot is greatly diminished. 

The renormalization constant $C$  is the probability of the conditioning event, i.e., the constant is itself also a PDF --- the PDF for the measurement of absolute phase which (like absolute time) is measureable in the fuzzy, albeit not in the sharp, sense. In considering a sequence of snapshot PDFs for various $\Phi_\Sigma$ we should dispense with the notion of absolute time marching along uniformly (as it has historically, as a parameter in quantum theory rather than an operator). Instead when we realize that the probability of the conditioning event is telling us that taking a snapshot at one time is not as probable as at some other time --- then we see more clearly how these snapshots turn into the quantum polarization ellipse. It is as if the polarization vectors spend more time in snapshots where $P(\Phi_\Sigma)$ is large and zip past (or even skip) snapshots taken at $\Phi_\Sigma$ of small (or even zero) $P(\Phi_\Sigma)$.  The time averaging inherent in the marginal POM, 
 naturally incorporates this weighting. 


\begin{figure}[h]
\centering 
\includegraphics[scale=.634]{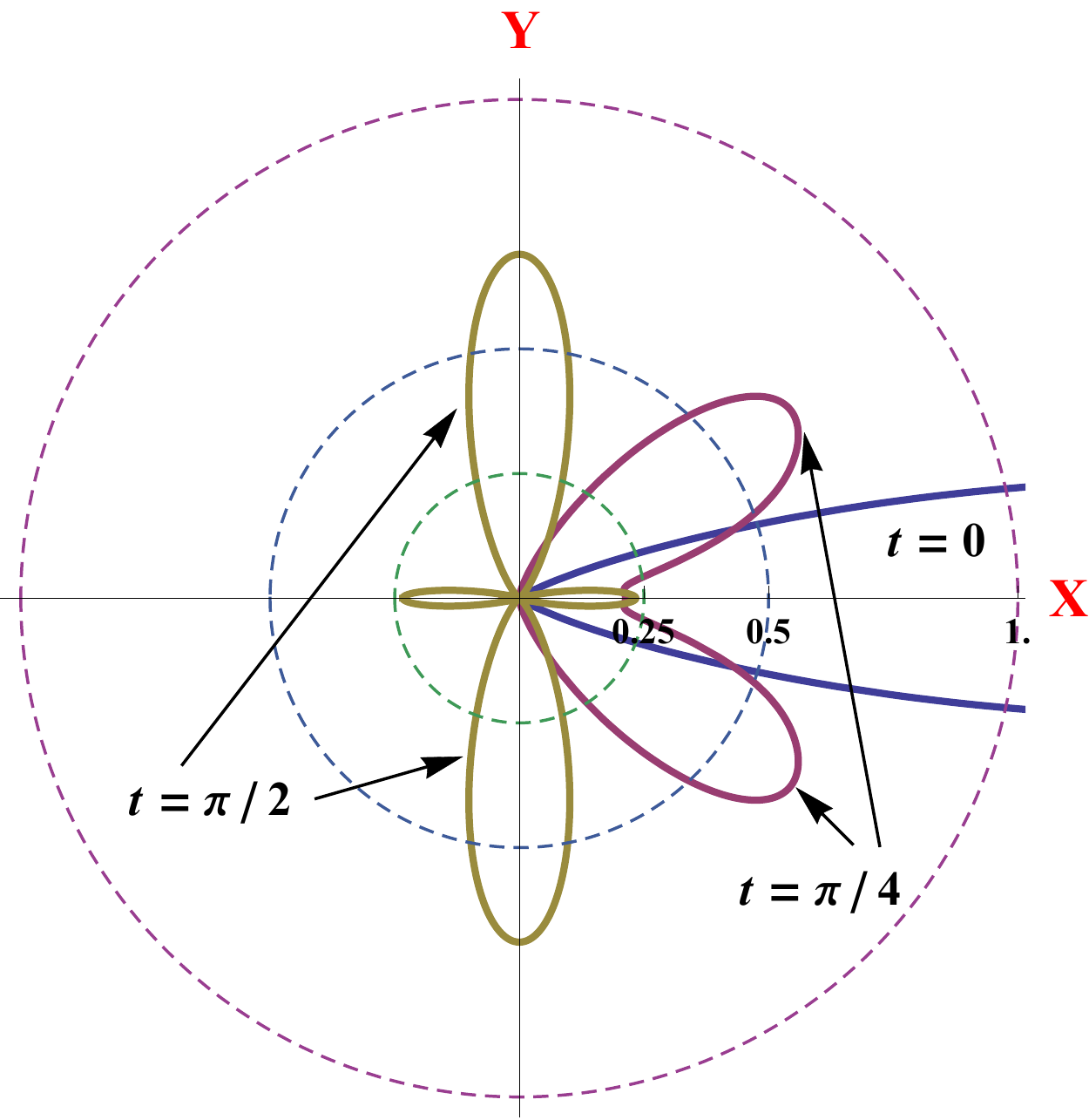}
\caption{Three Snapshot PDFs from Figure 7 --- Presented in a Polar Plot on a Linear Scale.}
\end{figure}

The quantum polarization ellipses for an x-polarized coherent state of $N = 1, 4$ and $9$ are 
presented on a dB scale in the polar plots of Figure 9. 
To better reveal the side-lobe structure it is preferable to plot the log of probability (rather than the probability itself) and to avoid a negative radius in a polar plot we add a scaling constant. In the graphics we arbitrarily scale each of these so that the peak of each of the PDFs corresponds to 60 dB (setting the peaks to some fixed reference level facilitates comparison of the underlying shapes). 
We see, for $N=9$, the probability of being ``up along the y-axis'' is almost 40 dB below the probability of being ``up along the x-axis.'' For $N=4$ we see this ratio is over 20 dB and for $N=1$ we are approaching the $N=0$ case of a uniform distribution (a circular polarization ellipse) since the vacuum state is rotationally invariant. 

\begin{figure}[h]
\centering 
\includegraphics[scale=.95]{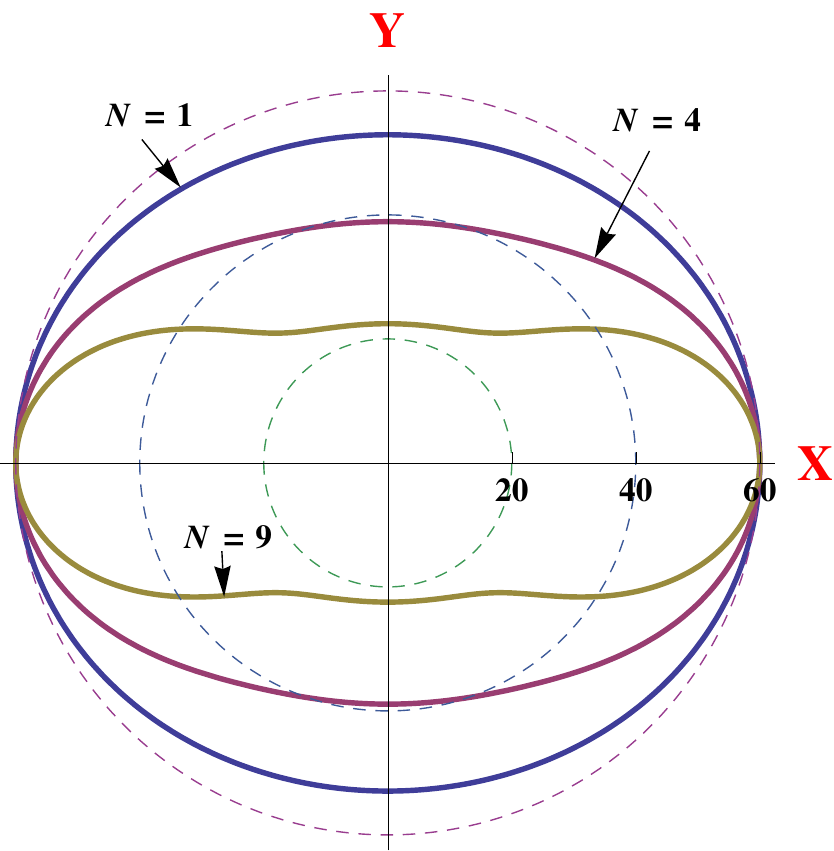}
\caption{Quantum Polarization Ellipses on a dB Scale for X-Polarized  Coherent States of $N= 1,4$ and $9$.}
\end{figure}

Of course an x-polarized coherent state is not exactly orthogonal to a y-polarized coherent state since they share the same vacuum state. But an x-polarized number state can also have an appreciable $P(\phi=\pi/2)$ although these are exactly orthogonal to any y-polarized number state (when that number is not zero). 
More generally, within the phase representation one can readily see that only terms of the form  \rm{Cos}\hspace{.5mm}$(m 
\phi)$ will arise for states of $\psi_{j,m}$ that are symmetric under $m\rightarrow -m$ [29 -- 31].
Any state for which the y-polarization mode is in the vacuum state will have this symmetry. For photons, $m$ will be odd when $j$ is odd ($m$ increments by two for photons, see Figure 3) and in that case the probability of $\phi=\pm \pi/2$ vanishes (for all $\Phi_\Sigma$). Thus, a single x-polarized photon will never have an electric field pointing along the y-axis (i.e., $P(\phi=\pm \pi/2)=0)$ when its quantum angle is measured, but a two-photon x-polarized field can! Indeed, any x-polarized field with 
non-vanishing components of even $j$ (such as a coherent state) will exhibit such behavior. In the classical limit $P(\phi=\pi/2)$ rapidly diminishes (even for $N$ only equal to $9$ its already less than $0.0002$) but for a weak coherent state of $N=1$ it is slightly over $6\%$. Since weak coherent states are sometimes used in polarization based quantum communication systems to mimic single-photon states, such an effect might merit consideration [32].

\section{concluding remarks}

The complementarity between time and energy, as well as between an angle and a component of angular momentum, was described at three layers of understanding complementarity in a more general context. The first layer, comprised of a simple Fourier transform of the complementary wavefunction, amounts to a non-projection-valued probability-operator measure and we elucidated ways in which these can be interpreted as fuzzy measurements. The phase of a single harmonic oscillator and the angle of a single particle are examples in which the limited dimensionality of the state space was shown to prevent wavefunction collapse in the phase or angle wavefunctions and further restrict the class of their realizable fuzzy measurement statistics. Such measurements can however be described via sets of commuting observables on a larger state space which includes an auxiliary system (which \textit{must} be a part of the apparatus which realizes the measurement) when the  auxiliary and original systems are not entangled prior to the measurement. 

Therein the auxiliary system is shown to function as a noise source and the fuzzy statistics manifest. Such extensions to larger state spaces are not unique and  are only intended to recover the fuzzy statistics. To go beyond these we must also extend the \textit{meaning} of what it is that we wish to measure. Clearly no general way of doing that can exist but using complementarity as a guide we were led to the conclusion that formally: only relative phase can be measured and the angle of a particle will require a field for its measurement. The meaning of relative phase gleaned at the second layer of understanding was achieved on an important subspace of the more general two-mode (two oscillator) state space. This subspace has some interesting physical properties (automatically entangling the two modes) which are also useful for quantum noise reduction and it is the space in which the celebrated N00N states reside. The phenomena of super-resolution is readily apparent in the quantum phase representation which also reveals that entanglement is not required and the N00N state performance can be identically reproduced on the single-mode space: simply notice that the periodicity of the magnitude square of a Fourier series is set by the minimal distance in $m$ between 
non-vanishing Fourier coefficients. 

In preparation for the final layer of understanding complementarity, Schwinger's harmonic oscillator model of angular momentum was modified to include the case of photons (instead of only the unrealizable fermionic primitives). Therein the quantum angle measurement (complementary to the measurement of a  component of angular momentum) was shown to be equivalent to the relative phase measurement between those two oscillators. 

The meaning of relative phase was finalized at the third layer. At the second layer there are an infinite number of subspaces that could be defined on a two-mode space wherein each value of photon number difference also corresponds to a unique value of number sum. The general theory (and final layer of understanding) on an unrestricted  two-mode space can cover all of those infinite number of possible measurements and it demonstrates two reasonable ways of dealing with the degeneracy in number sum. These also correspond to two reasonable ways of eliminating absolute time (which is measureable in a fuzzy albeit not in a sharp sense)  in order to define a direct measurement of the relative phase: a conditional measurement which takes a snapshot in absolute time (corresponding to adding probability amplitudes); and a marginal measurement which takes an average in absolute time (corresponding to adding probabilities). The sense in which distinguishability is a ``matter of how long we look'' was discussed and the meaning of the general theory was illustrated by taking the two oscillators to model the right and left circularly polarized modes of an electromagnetic plane wave so that the conditional  measurement reveals a snapshot of the angular distribution of the electric field vector and the marginal measurement corresponds to a quantum version of the polarization ellipse. 

The quantum angle representation demonstrated that any excitation of an odd number of x-polarized photons will \textit{never} have an angle in correspondence with the y-axis; but that of  an even number of x-polarized photons \textit{always} can! The behavior of an x-polarized coherent state was examined and the snapshot angular distributions were seen to evolve into two counter-rotating peaks resulting in considerable correspondence with the y-axis (particularly for weaker coherent states) at the time for which a classical linear polarization vector would shrink to zero length. 
 Such an effect could be of significance for polarization based quantum communication systems 
 since weak coherent states are sometimes used to mimic single-photon states.
We also demonstrated how the probability distribution of absolute time (now treated as a measurable quantity, rather than just a parameter) has an influence on how these snapshot angular distributions trace out and evolve into the quantum polarization ellipse.


\vspace{.15in}
\hspace{-0.13in}{\textbf{Appendix 1: Alternate Path to the Single-Mode Statistics}}
\vspace{.15in}

Concurrent to (and independent of) the development of the continuous single-mode phase representation, an alternate method for obtaining the single oscillator statistics was derived by Pegg and Barnett [9], [10]. Their approach requires the truncation of the infinite dimensional state space of a harmonic oscillator to one of finite but arbitrarily large dimension. This subspace, denoted ${\mathcal{H}^T}(s)$, is spanned by the number-kets $\{ | n \rangle : 0 \le n \le s \}$. Furthermore, their formalism relies on an ordering of terms, in a polar decomposition of the annihilation operator, which is akin to Dirac's ordering [1] rather than Susskind and Glogower's [2].
This ordering permits them to impose an ``additional condition,'' which specifies the action  on the vacuum state --- whereas the action of the SG operator
on the vacuum state is uniquely determined by that of $\hat{a}$, as mentioned.

For any number state $|n\rangle$, with $n  \in  \{1,2, . . . \ s \}$ but $n \ \not= \ 0$, the unitary power series of Pegg and Barnett's operator, exp$(i \hat{\phi}_{PB})$ defined on ${\mathcal{H}^T}(s)$, is a lowering operator. The action of this operator on the vacuum state is then defined to be a ``wrap-around'' term:
\begin{equation}
{e^{i\hat{\phi}}}_{PB} | 0 \rangle \equiv e^{i(s+1)\theta_0} | s \rangle
\end{equation}
where $\theta_0$ is the location of the branch cut for phase (which is $-\pi$ in our formalism). This cyclic behavior, possible only in a truncated space such as ${\mathcal{H}}^T (s)$, is essential for their definition of an Hermitian phase operator on ${\mathcal{H}^T}(s)$. Unitarity is accomplished by not having to ``stop'' at the vacuum state, but the ``wrap-around to the top of the stack'' term  causes the discrete-phase wavefunctions to be   
complementary to a periodically replicated (and hence truncated) version of the $\{ \psi_n \}$.

This operator can be expressed in terms of an orthogonal subset, $\{ | \theta_m \rangle \}$, of truncated phase-kets as
\begin{equation}
\hat{\phi}_{PB} = \sum_{m=0}^{s} \theta_m | \theta_m \rangle \langle \theta_m |,
\end{equation}
where
\begin{equation}
|\theta_m \rangle \equiv (s+1)^{-1/2} \sum_{n=0}^{s} e^{in\theta_m} | n \rangle \ \ \text{and} \ \ \theta_m \equiv \theta_0 + 2\pi (\frac{m}{s+1}).
\end{equation}
A measurement of $\hat{\phi}_{PB}$ on ${{\mathcal{H}}^T}(s)$ will yield one of its discrete eigenvalues, $\theta_m$, which are rational multiples of $2\pi$ plus $\theta_0$, with probability:
\begin{equation}
P_r (\theta_m) = {\mid}\langle \theta_m | \psi \rangle^T {\mid}^2 ,
\end{equation}
where $|\psi {\rangle}^T$ is a truncated state, $\sum_{n=0}^{s} \psi_n | n \rangle$, on ${\mathcal{H}^T}(s)$.

As the truncation point, $s$, goes to infinity, Pegg and Barnett's discrete phase eigenspectra converge to our phase continuum (in as much as rational numbers can converge to real numbers) i.e., their probability mass function (72) converges in distribution to our probability density. It is this $s \ \rightarrow \ \infty$ limit in which they argue that the discrete phase statistics make physical sense. It should be emphasized that although Pegg and Barnett indeed have an Hermitian phase operator on ${\mathcal{H}^T}(s)$ --- with $s$ finite; when $s \ \rightarrow \ \infty$ this is an alternative means of calculating the fuzzy  single-mode continuous phase measurement statistics  via a limiting procedure. If $s$ is left to be finite then their discrete phase statistics correspond to a discrete Fourier transform --- complementary to a truncated (and periodically replicated) version of the $\{ \psi_n \}$. In the limit $s \ \rightarrow \ \infty$ these converge (in distribution) to the continuous phase statistics which we can obtain directly via the Fourier transform of the actual $\{ \psi_n \}$. Note that by accepting the non-Hermitian nature of phase, we were led to the necessity of a larger state space which emphasizes the existence of an auxiliary noise source.

\vspace{.15in}
\hspace{-0.13in}{\textbf{Appendix 2: General Theory of Complementarity}}
\vspace{.4mm}

Herein we analyze complementarity at the first (fuzzy) layer, which is general in the sense that it can cover an angle, phase, or time (in the case of a time independent Hamiltonian) when there is \textit{no degeneracy in the complementary eigenspectra}. To cover  cases of degenerate eigenspectra one must use procedures akin to (61) and/or (64). 
At layer one, the proof can be made to follow (almost verbatim) the case which stems from linear momentum being the generator of translations in space [23]. 

Postulate the existence of a set of eigenstates  $| \chi\rangle$ which resolve the identity operator so that these correspond to a realizable measurement. Furthermore, let the eigenvalues $\chi$ be continuous, non-degenerate and real valued. Denote the operator which effects an infinitesimal translation in $\chi$ by $\hat{T}(d\chi):$
\begin{equation}
\hat{T}  (d\chi) | \chi \rangle = | \chi + d\chi \rangle.
\end{equation}
The following physically reasonable properties:
\begin{equation}
\hat{T}^{\dagger}(d\chi) \ \hat{T}(d\chi)=\hat{I} 
\end{equation}
\begin{equation}
\hat{T} (d\chi_1) \ \hat{T}  (d\chi_2) = \hat{T} (d\chi_1 + d\chi_2)
\end{equation}
\begin{equation}
\hat{T} (-d\chi) = \hat{T}^{-1} (d\chi) \ \ \text{and} \ \ \lim_{d\chi\to 0} \hat{T}  (d\chi) = \hat{I}
\end{equation}
are obtained (for {\it{infinitesimal}} d$\chi$) when
\begin{equation}
\hat{T} (d\chi) = \hat{I} - i \hat{G} d\chi,
\end{equation}
where $\hat{G}$ is said to be the generator of translations in $\chi$. An example of this is when $\chi$ is space and $\hat{G}$ is proportional to linear momentum
with Planck's constant incorporated into its definition so that (77) is dimensionless.
To progress from this to the Fourier transform between representations (and hence achieve complementarity) we will assume that $\hat{G}$ is Hermitian (self-adjoint) so that its eigenspectra are real. 
 If however, we try to take $\hat{G}$ to be an angle operator, for example, generating finite (not infinitesimal) translations in $m$ then complications arise (differentiation not being defined for a discrete parameter being the least of them) 
and indeed these complications are trying to tell us something ``is wrong'' here which then forces us to a higher dimensional state space in order to achieve a complete description (i.e., a sharp measurement) in terms of sets of commuting observables. We can however remain at the first layer and obtain an incomplete description (i.e., a fuzzy measurement) if instead we take an angle to be $\chi$. Note however that in (73) there can be no ``stopping''  --- as in the sense of the SG operator stopping at the vacuum. I.e., the eigenspectra of $\chi$ must range from $-\infty$ to  $\infty$ 
else (73) cannot hold $\forall d\chi$ and $\forall \chi$ --- which would preclude the definition of a derivative in what follows, i.e., (79).
\textit{Later}, when we find the angle distribution to be periodic mod $2\pi$ (although clearly fermions can exhibit mod $4\pi$ behavior the observation of such requires their interference with another system) \textit{then} we can restrict our attention to one of these identically distributed $2\pi$ intervals. 

To be sure, avoidance of stopping is what leads us to (at the second layer) extend the SG operator to one on $\mathcal{H}'$, or to define a lowering of $m$ on a field/particle system. Rather than taking a phase operator to generate translations in photon number we can take $\hat{n}$ as the generator of translations in $\phi$ (as indeed it is already accepted that the Hermitian Hamiltonian generates translations in time, and $\hat{J_{z}}$ generates translations in the angle about the z-axis, etc.) and therein we can remain at the first layer and simply justify the Fourier transform that leads to the fuzzy complementary measurement statistics.  

Let $\psi(\chi) \equiv \langle\chi|\psi\rangle$, from (73) and (77)  we have 
\begin{equation}
\langle \chi | \hat{T}^{\dagger}(d\chi) | \psi\rangle = \psi(\chi+d \chi) = \psi(\chi) + i \langle \chi | \hat{G}^{\dagger}
 | \psi \rangle d\chi
\end{equation}
so that
\begin{equation}
\frac{d \psi}{d\chi}= i \langle \chi | \hat{G} | \psi \rangle \longrightarrow
\frac{d }{d\chi} \langle \chi  | G \rangle = i \hspace{.5mm}  G \ \langle \chi  | G \rangle
\end{equation}
when we take $|\psi\rangle$ to be an eigenket of $\hat{G}$ (and we used $\hat{G}^{\dagger} = \hat{G}$.) The solution to this differential equation is  $\langle \chi |G\rangle = N e^{iG\chi}$, the kernel of the Fourier transform, where N is a normalization constant. The presumed completeness of the $| \chi \rangle$ then leads to 
\begin{equation}
\langle G|\psi \rangle =  \int d\chi \ \langle G|\chi \rangle \langle \chi |\psi \rangle
\end{equation}
 i.e., with $\psi_{G}(G) \equiv \langle G|\psi \rangle$ we have 
\begin{equation}
\psi_{G}(G) = N \int d\chi \ e^{-iG\chi} \ \psi (\chi)
\end{equation}
from which the inverse Fourier transform follows.
Then Rayleigh's energy theorem (or Parseval's power theorem) [15] proves that a normalized distribution in one domain will have a complementary distribution which is also normalized (in the complementary domain). Thus, if the eigenkets of the generator, $|G\rangle$, are complete in some space then the eigenkets $|\chi\rangle$ are complete in that same space, proving the assumption, which concludes the proof (at the first layer).

If the original distribution, i.e., that of the Hermitian generator of translations, such as a Hamiltonian,  is continuous and aperiodic (i.e., not periodic) then the complementary distribution is also continuous and aperiodic, and the two wavefunctions are related by the Fourier integral transform. If the original distribution is ``rationally-discrete'' (i.e., in correspondence with numbers whose ratios are rational numbers) and aperiodic, then the complementary distribution will be continuous and periodic (the Fourier series transform relationship).  If the original distribution is aperiodic and in correspondence with a discrete set of numbers whose ratios are \textit{not} rational numbers, then the complementary distribution will be continuous and ``quasi-periodic.'' Lastly, if the original distribution is discrete and periodic, then the complementary distribution will be also (the discrete Fourier transform). Thus, for any system with a rationally-discrete energy spectrum the temporal distribution will be periodic. Likewise, the only system which exhibits truly discrete temporal behavior is one in which the energy distribution is truly periodic. Similarly, the quantization of angular momentum (projected onto an axis) is the simple and immediate consequence of the periodicity of the angle about that axis.



%
%

%



\begin{thebibliography}{99}
\bibitem{3} Formally a Hilbert space corresponds to a non-denumerably infinite vector space (in correspondence with the set of real numbers) rather than a denumerably infinite space (e.g., in correspondence with the set of  positive semi-definite integers). The later is occasionally referred to as a rigged Hilbert space; but  it has become common practice in the literature to refer to even a finite dimensional set of vectors (e.g., the eigenkets of the z-component of spin for a particle of finite total spin) as a Hilbert space. Since the dimensionality of the space in what follows is more than a matter of semantics we shall refer to these as a state space and reserve the phrase ``Hilbert space'' to those spanned by eigenkets of unbounded real eigenvalues.
\bibitem{16}
The notion that quantum measurements correspond to Hermitian operators (which have real eigenvalues) originally  stemmed from the notion that all measurements yield real numbers. Yet a complex number is merely a way of representing two real numbers simultaneously; and no one would argue that we could not measure the operator $\hat{o} = \hat{x} + i \hat{y}$, where $\hat{x}$ and $\hat{y}$  are the commuting (hence simultaneously measureable) $x$ and $y$ position operators. Therein we have simply labeled the y-axis as imaginary and reported the outcome as a complex number, rather than a vector. Thus we might restrict the class of realizable measurements  to sets of operators which commute with their adjoint, rather than to operators which are equal to their adjoint, since then their real and imaginary components (which are Hermitian by construction) will also commute. We have referred to such measurements as sharp and for these we have a complete  description of their measurement in terms of sets of commuting Hermitian operators.  Therein all of the conventional descriptions of quantum theory hold and the predilection for Hermitian operators rightfully stems from the fact that their eigenkets are orthogonal (hence providing the sharpness which permits wavefunction collapse).
\bibitem{18} In general these could be quantum fluctuations, but in our quantum optics examples these are zero-point fluctuations since therein the auxiliary system  is a mode which is in the vacuum state prior to the measurement.
\bibitem{25}   Equivalently,  if we wish to describe the angle measurement in terms of operators rather than wavefunctions, a field is required so that we can generate translations in an unbounded m-space (as exemplified by the arrows in Fig. 3 within the field/particle system). Only then can 
such translations be both-sided unitary and thereby define an Hermitian angle operator. If we instead introduce a wrap-around term (depicted by the semi-circle) and stay on the finite dimensional  state space of the particle then we lose complementarity.
\bibitem{29} Such states are also symmetric under $\phi \rightarrow -\phi$, the unitary time-reversal eigenkets, which constitute a generalization of ``x-polarized'' states, in which the y-polarization mode need not be in the vacuum state.
\bibitem{32}  An apparatus which realizes the quantum angle measurement has yet to be realized so one must carefully consider whatever apparatus (polarization beam splitter, etc.) is actually used. Nevertheless, the quantum phase and angle representations are so fundamental that their predictions are often discernable in other quantum measurements.
\end{thebibliography}
%

\vspace{2mm}
\hspace{-0.13in}{\textbf{Acknowledgments}}
\vspace{2mm}

\noindent The author would like to acknowledge many useful discussions with J. H. Shaprio on quantum measurements and support from NASA
EPSCoR (NNX13AB14A) and the Louisiana Board of Regents.

\end{document}